\documentclass[]{aa}  
\usepackage{graphicx}
\usepackage[authoryear]{natbib}
\bibpunct{(}{)}{;}{a}{}{,}
\usepackage{txfonts}
\begin{document}
   \title{The population of OB supergiants in the starburst
   cluster Westerlund~1\fnmsep\thanks{Based on observations collected
   at the European Southern Observatory, Paranal, Chile (ESO 71.D-0151, 073.D-0327, 081.D-0324)}} 


   \author{I.~Negueruela
          \inst{1}
          \and
          J.~S.~Clark\inst{2}
	  \and
          B.~W.~Ritchie\inst{2,3}
          }

   \offprints{I.~Negueruela}

   \institute{Departamento de F\'{\i}sica, Ingenier\'{\i}a de Sistemas y
  Teor\'{\i}a de la Se\~{n}al, Universidad de Alicante, Apdo. 99,
  E03080 Alicante, Spain\\
              \email{ignacio@dfists.ua.es}
         \and
	 Department of Physics and Astronomy, The Open University,
  Walton Hall, Milton Keynes MK7 6AA, United Kingdom
\and
IBM United Kingdom Laboratories,
Hursley Park, Winchester, SO21 2JN, UK
             }

   \date{Received }
\titlerunning{OB supergiants in Westerlund 1}

  \abstract
   {After leaving the main sequence, massive stars undergo complex evolution, which is still poorly understood. With a population of hundreds of OB
  stars, the starburst cluster Westerlund~1 offers an unparallelled environment to study their evolutionary tracks. }
   {We characterise a large sample of evolved OB stars in the cluster, with the aim of determining cluster parameters and place stars in an evolutionary sequence.} 
   {We used the FORS2 instrument on the VLT to obtain intermediate-resolution spectroscopy over the range 
  5800--9000\AA\ of about a hundred stars selected as likely members
  of the cluster based on their photometry. We developed criteria for their spectral classification using only spectral features in the range observed.  We discuss these criteria, useful for spectral classification of early-type stars in the GAIA spectral region, in the appendix. Using these criteria, we obtain spectral classifications, probably accurate to one subtype, for 57 objects, most of which had no previous classification or a generic classification.}
   {We identify more than 50 objects as OB supergiants. We find almost 30 luminous early-B supergiants and a number of less luminous late-O supergiants. In addition, we find a few mid B supergiants with very high luminosity, some of them displaying signs of heavy mass loss. All these stars form a sequence compatible with theoretical evolutionary tracks. In addition, two early B supergiants also show indication of heavy mass loss and may represent the evolutionary phase immediately prior to the Wolf-Rayet stage. We investigate cluster properties using the spectral types and existing photometry. We find that the reddening law to the cluster does not deviate strongly from standard, even though extinction is quite variable, with an average value $A_{V}=10.8$.  Though evolutionary tracks for high-mass stars are subject to large uncertainties, our data support an age of $\ga5$~Myr and a distance $d\approx5$~kpc for Westerlund~1.}
   {The spectral types observed are compatible with a single burst of star formation (the age range is very unlikely to be $>1$~Myr). Westerlund~1 shows its potentiality as a laboratory for massive star evolution, which can be fulfilled by detailed study of the population presented here.}


   \maketitle
%

\section{Introduction}

 Massive stars play a crucial role in the dynamical
and chemical evolution of galaxies, providing a major source of 
ionising UV radiation, mechanical energy, and chemical enrichment
\citep[e.g.,][]{mas03,lanza}. As they evolve from the main
sequence (MS) towards the Wolf-Rayet (WR) phase, massive stars
must shed most of their outer layers.  Models
predict that massive stars will evolve redwards at approximately
constant $L_{{\rm bol}}$, but there is an apparent dearth of yellow
supergiants with  $L_{{\rm bol}}$ comparable to the brightest
hot stars. The lack of yellow hypergiants above a certain luminosity defines the observational Humphreys-Davidson
(HD) limit, which seems to imply that stars hit some kind of instability
when they reach this area of the HR diagram, and then lose mass at a
formidable rate. This phase of enhanced mass loss is generally
identified with the luminous blue variable (LBV) stage, but there is
no clear understanding of the actual evolution through the HR diagram
of massive stars during the H-shell burning phase \citep{hd94,vGen01}.

\defcitealias{main}{C05}

There is general agreement that the WR stage corresponds to the
  He-core burning phase, and the fact that WR stars are very hot
  implies that massive stars do actually loop bluewards. But there is
  a complex zoo of transitional objects, comprising blue supergiants (BSGs), red supergiants (RSGs), 
  yellow hypergiants (YHGs), LBVs and OBfpe/WNVL stars, whose
  identification with any particular evolutionary phase is a matter of
  guesswork. Understanding this evolution is, however, crucial because
  the mass loss during this phase completely determines the
  contribution that the star will make to the chemistry of the ISM and
  even the sort of post-supernova remnant it will leave.

Unfortunately, massive stars are scarce and, as this phase is very short on
evolutionary terms, examples of massive stars in transition are
rare. For most of them, 
distances are unknown and so luminosities can be determined at best to
order-of-magnitude accuracy, with the uncertainty feeding through to other parameters ($M_{*}$, $R_{*}$). As a consequence, the difficulty in placing these objects within an evolutionary sequence is obvious.

The starburst cluster Westerlund~1 (Wd~1) may represent a unique laboratory for addressing these issues. With a mass $\sim10^{5}\:M_{\sun}$ \citep[][]{brandner,mt07}, it is young enough to contain a large number of OB stars, but old enough to contain stars in all the evolutionary stages. Large populations of WR stars and transitional massive stars
\citep[][henceforth C05]{wrs,main} have been identified. In this paper, we set their evolutionary context by characterising the population of moderately evolved massive stars in the cluster, i.e., objects that have started to move away from the MS, but have not yet reached the region of instability. With this information, we are able to constrain the cluster parameters and check the agreement with evolutionary tracks, which supports the idea of an (approximately) single-age population.


\section{Observations}

 Observations of stars in Wd~1 were carried out on the nights of 2004
 June 12th and 13th using the spectro-imager FORS2 on Unit~1 of the
 VLT (Antu) in three different modes: longslit, multi-object
 spectroscopy with masks (MXU) and multi-object spectroscopy with
 movable slitlets (MOS). We used grisms G1200R and G1028z, which provide the highest dispersions, to obtain
 intermediate-resolution spectroscopy. G1200R has a nominal dispersion of 0.38 \AA/pixel over the spectral range 5750--7310\AA. G1028z has a nominal dispersion of 0.42\AA/pixel over the spectral range 7730--9480\AA. Together, they cover most of the red and nearest infrared range.

 For the longslit mode, we used a $0\farcs3$ slit to obtain the highest resolution available. Measurements of arc lines give FWHM of $\sim$0.9\AA\ with the G1200R, which results in $R$ slightly above $\sim7000$. With the G1028z, we measure FWHM$\sim1.3$\AA, implying a resolution slightly below $\sim7000$.
For the multi-object modes, we used $1\arcsec$ slits ($10\arcsec$ long in
 the MOS mode and of variable -- $<10\arcsec$ -- length in the MXU
 mode). This results in a resolving power $R=2200$ in the red range (the exact
 wavelength coverage depends on the position in the CCD) and $R=2500$ in
 the near-IR range.

%
   \begin{figure}[ht]
   \centering
\resizebox{\columnwidth}{!}{\includegraphics[angle=-90]{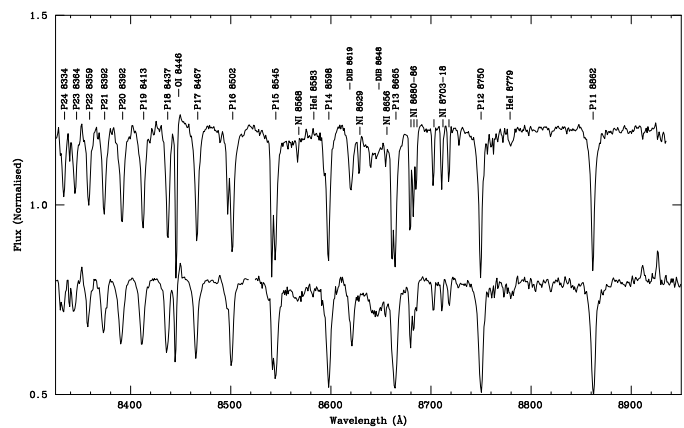}}
\caption{$I$-band spectra of two of the visually brightest B
  supergiants in Wd~1, tentatively classified as hypergiants in \citetalias{main}. 
The top spectrum corresponds to W42a, which we
     classify as B9\,Ia$^{(+)}$, based on the features presented
       here. Note that the \ion{Ca}{ii} triplet lines are already
       comparable in strength to Pa 13, 15 \& 16. The bottom spectrum
       corresponds to W33, B5\,Ia$^{(+)}$. Note the prominent P-Cygni
        profile in \ion{O}{i}~8446\AA\ and the emission features at
         8913 and 8927\AA, seen in the spectra of all the mid-B
         supergiants in Wd~1.}
              \label{hyperblues}
    \end{figure}
%

 We also used the MOS mode with grism G150I to obtain low-resolution
 spectroscopy of stars in the field. This mode gives a resolving power $R=260$
 over a range $\sim6000$--11000\AA\ (the exact range depending again on the
 position within the CCD). This mode was used mainly for
 Wolf-Rayet stars, but other objects were also observed.
 
 The field of view of FORS2 is $6\farcm8\times6\farcm8$ . Within the central
 $5\farcm0\times5\farcm0$ field of view, we selected our targets
 from the list of likely members of \citetalias{main}. For the external
 regions, targets were selected at random amongst relatively bright
 stars. In total, we took three MXU and one MOS mask with both G1200R and
 G1028z grisms, two further MXU masks with only the G1200R (these were
 aimed at relatively faint objects, which were expected to be OB stars
 near the MS and so not to have strong features in the range covered
 by G1028z) and three MOS masks with G150I.
 This resulted in $\sim 120$ stars observed with G1200R, $\sim75$
 stars observed with G1028z and $\sim 50$ stars observed with
 G150I. More than 90\% of the spectra turned out to correspond to OB
 stars, and hence cluster members.

In this paper, we restrict our analysis to the brightest OB members of the cluster, leaving analysis of the unevolved population of OB stars for a future work, where our FORS2 data will be combined with FLAMES and ISAAC observations. We complement our dataset with observations of a few additional bright OB stars which could not be observed during the FORS2 run. A few stars (W5, W7 \& W52) were observed with the ESO Multi-Mode
Instrument (EMMI) on the New 
Technology Telescope (NTT) at the La Silla Observatory (Chile) on the nights of June 5--8 2003, using gratings \#6 \& \#7, giving a coverage similar to that of the FORS observations at comparable resolution (see \citealt{clark09} for details). Some other stars were observed with the NTT and EMMI on the 
nights of 15\,--\,17 February 2006. On this occasion, we used grism \#6, which covers the range 5800--8650\AA\ at a resolution $R\sim1500$, though several objects are only detected from $\ga6500$\AA.

For completeness, we include a few other objects which have been observed only in the $I$ band with the multi-object fibre spectrograph FLAMES, on the VLT. Details of those observations are reported in \citet{ritchie09a}. The spectra cover the range 8484--9000\AA, with a resolving power $R\approx16,000$.

%
   \begin{figure*}[ht]
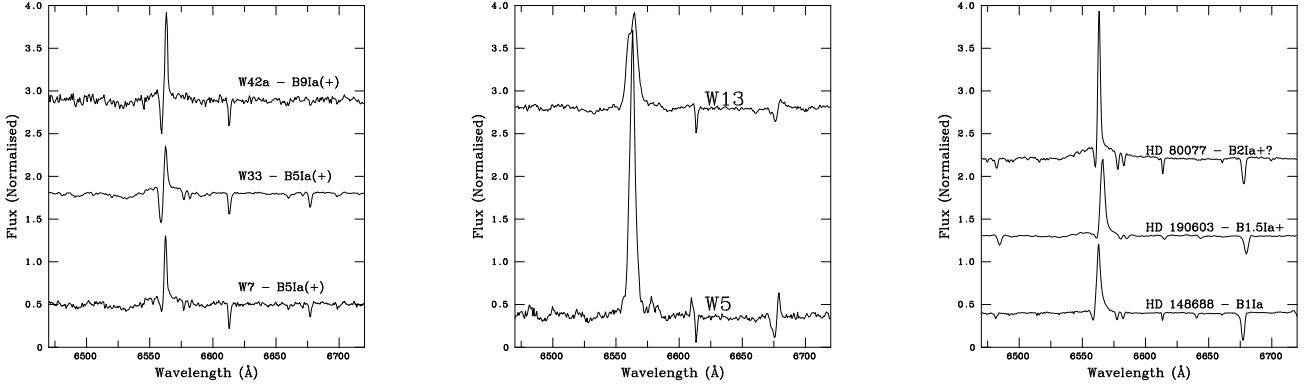

   \centering
\resizebox{\textwidth}{!}{
\includegraphics{panel1.eps}
\includegraphics{panel2.eps}
\includegraphics{panel3.eps}}
\caption{H$\alpha$ spectra of the most luminous B-type stars in Wd~1
  and some comparison stars. The left panel shows the spectra of the
  three luminous B-type objects tentatively classified as hypergiants
in \citetalias{main}. The middle panel shows two objects with
spectral types around B0 with evidence for heavy mass loss. The right panel shows, as
comparison, the spectra of the hypergiant candidate HD~80077, the
hypergiant HD~190603 and the supergiant HD~148688, which has the
strongest emission feature amongst all the objects examined in
Appendix~A. All the spectra have had the continuum normalised and are offset for display.}
              \label{fighyperalpha}
    \end{figure*}

\section{Results}
We have selected for our sample those bright stars not listed as yellow or red supergiants \citepalias{main} or Wolf-Rayet stars \citep{crowther}. The sample consists of 57 stars spectroscopically identified as luminous OB cluster
 members. They are listed in Table~1. 


\begin{figure*}[ht]
\centering
\resizebox{\textwidth}{!}{\includegraphics[angle=-90]{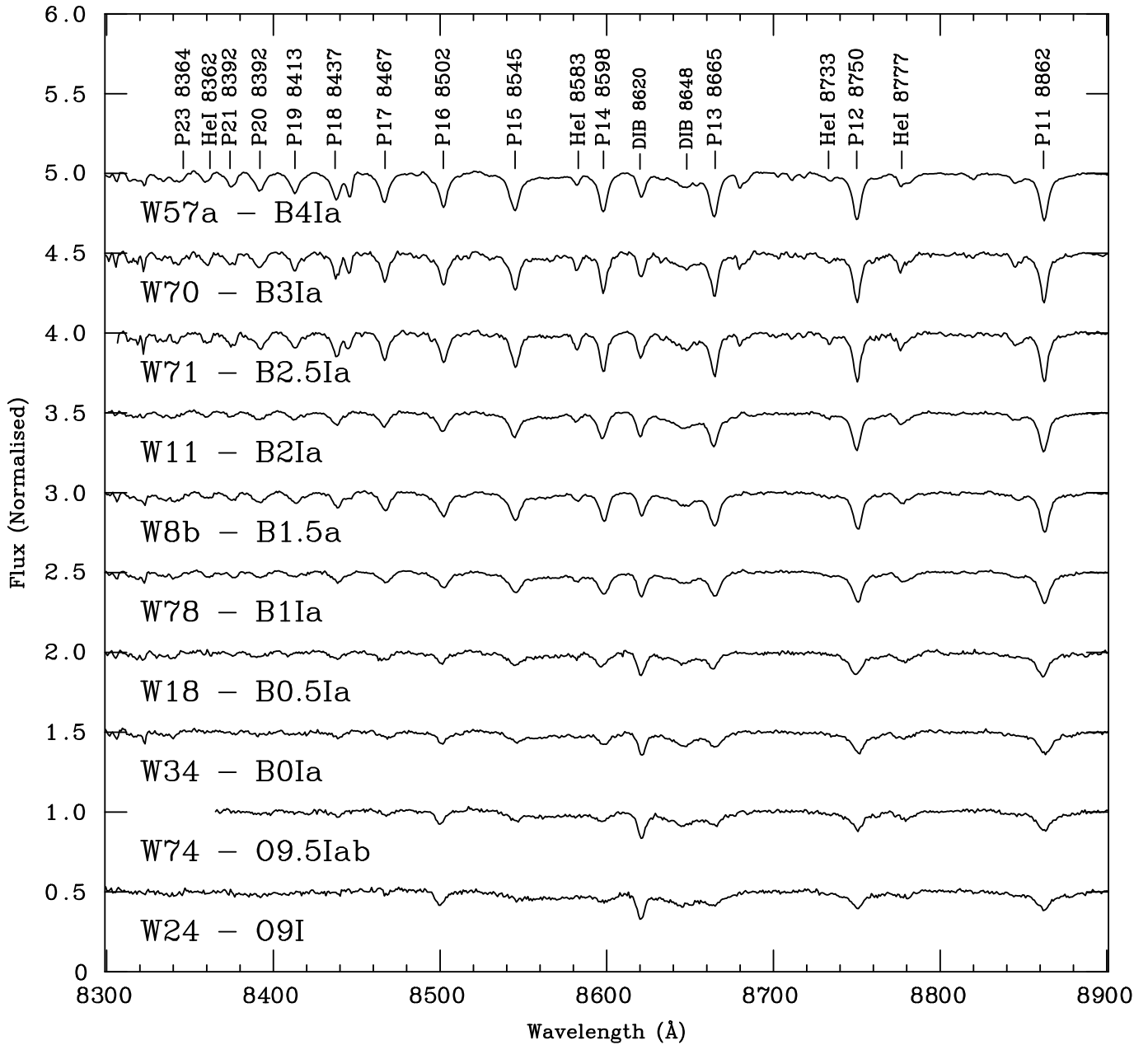}} 
 \caption{Sequence of $I$-band spectra of blue supergiants in Wd~1,
   showing the evolution of the main features. Note the disappearance
   of N\,{\sc i} features between Pa~12 and Pa~13 and O\,{\sc
     i}~8446\AA\ around B2 (compare to Fig.~\ref{hyperblues}, where lines are identified) and the 
   development of Pa~16, which shows the presence of a strong C\,{\sc
     iii} line for stars B0 and earlier..}
   \label{isequence}
\end{figure*}

 Wd~1 is too obscured to allow the acquisition of classification
 spectra. Even with 8-m class telescopes, moderate SNR cannot be
 achieved at wavelengths shorter than 5500\AA\ with reasonable
 exposure times. We have been forced to rely on the data available,
 covering the range between $\sim$6000\AA\ and 9000\AA. Criteria for
 classification in this range were first explored in
 \citetalias{main}, and are analysed and developed in
 Appendix~A.

As demonstrated in the appendix (see also Appendix~A of
\citetalias{main}), the classification of supergiants of spectral types
  later than B2 can be achieved with rather high accuracy, because of
  the presence of abundant metallic lines. For earlier spectral types,
  classification criteria are scarcer and most of them are
sensitive to both spectral type and luminosity. In most cases, by using several
of them at the same time, we can narrow the spectral range for a given
spectrum generally down to half a spectral subtype. This is achieved here
by combining the red and $I$-band spectra.

There are three very luminous B supergiants in Wd~1, which were
tentatively classified as hypergiants in \citetalias{main}. The
$I$-band spectra of W33 and W42a are displayed in
Fig.~\ref{hyperblues}. Based on them, they can be confidently
classified as B5\,Ia$^{(+)}$ and B9\,Ia$^{(+)}$. W7 is very similar to
    W33, and was also classified B5\,Ia$^{(+)}$\fnmsep\footnote{Note that \citet{clark09} observe pulsational variability in these objects and speculate that they may induce variations of up to 1 spectral subtype}. These objects display weak P-Cygni profiles in the \ion{O}{i}~8446\AA\ line, and much
     stronger P-Cygni profiles in H$\alpha$ (see
     Fig.~\ref{fighyperalpha}). Another interesting feature in the
     red spectra of W7 and W33 is the \ion{C}{ii}~6578, 6582\AA\
     doublet. These lines are hardly seen in MS B-type stars,
but appear weakly in B2--B5 stars of moderate luminosity and can be
prominent in
supergiants. Its maximum strength occurs around B3
\citep{wal80}. 

Amongst Wd1 stars, the strongest \ion{C}{ii} is seen in
the spectra of W70 and W71, both of which were classified in
\citetalias{main} as B3\,Ia based on their 
$I$-band features (strong \ion{O}{i}~8446\AA\ line by the side of
Pa~16, presence of \ion{N}{i} lines around Pa~11). Another object with
prominent \ion{C}{ii} doublet is W57a. From the ratio of
\ion{O}{i}~8446\AA\ to Pa~16 and the intensity of the \ion{He}{i}
lines, this object is either later or more luminous than W70 (Fig.~\ref{isequence}). From the
colours and magnitudes of these stars, the last possibility seems
unlikely (all three have about the same magnitude in $I$, while W57a
is less reddened). This is supported by the weaker \ion{C}{ii}
doublet. W57a shows distinctive \ion{Si}{ii}~6347 \& 6371\AA, as well as
\ion{Ne}{i}~6402\AA. These lines are also strong in W70, but weaker in
W71. Therefore, we take B4\,Ia for W57a and B3\,Ia for W70. W71
is slightly earlier, and we classify it B2.5\,Ia.

\begin{figure}[htb]
\centering
\resizebox{\columnwidth}{!}{\includegraphics[angle=-90]{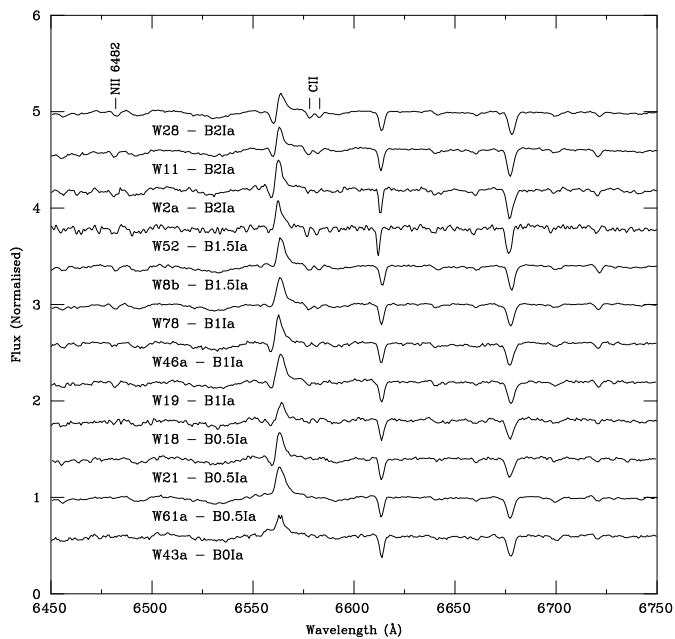}} 
 \caption{Sequence of red spectra of blue supergiants in Wd~1. The
   shape of H$\alpha$ changes from a P-Cygni profile to pure emission
   around spectral type B0.5. The N\,{\sc ii}~6482\AA\ line is only
   prominent in Ia supergiants and its strength peaks sharply at B2. }
   \label{rsequence}
\end{figure}

Then we have W2a and W28, presenting moderately strong
\ion{C}{ii}~6578,6582\AA\ (Fig.~\ref{rsequence}) and still a conspicuous \ion{O}{i}~7774\AA\ triplet, a strong feature in the later supergiants.
The \ion{O}{i}~8446\AA\ line is seen just as an inflection on the side
of Pa~18 (Fig.~\ref{isequence}). Both display strong P-Cygni profiles in H$\alpha$
and strong \ion{N}{ii}~6482\AA. The presence of this line is
considered a signature of Ia class supergiants by \citet{lenn93}. We classify
these objects B2\,Ia.  W11 is very similar and receives the same classification. W23a has a very similar $I$-band spectrum. It
also shows \ion{N}{ii}~6482\AA. However, \ion{O}{i}~7774\AA\ is
missing, and the \ion{C}{ii} doublet is very weak. These features are
difficult to reconcile with any spectral type. As this star is very
bright, we suspect that the spectrum may include an earlier type
companion. W8b and W52 look slightly earlier than this previous group, but still have strong \ion{C}{ii}~6578,6582\AA\ and perhaps \ion{O}{i}~8446\AA\
on the wing of Pa~18. We
classify them as B1.5\,Ia.

W78, still showing
moderately strong \ion{C}{ii} doublet lines, has weaker
\ion{O}{i}~7774\AA. The absorption trough in the P-Cygni profile of
H$\alpha$ has almost disappeared in this object. The spectrum of W78
extends further bluewards than most others, and we can identify a very strong
\ion{Si}{iii}~5740\AA\ line, confirming a spectral type not far away
from B1. Similar features are seen in W46a, W43b and W19. Though there
is some evidence that these last three may be very slightly earlier,
we classify all these stars as B1\,Ia.

W18 and W21 show inconspicuous \ion{C}{ii} absorption lines and
broader Paschen lines, though we can still see them up to Pa~20. The
\ion{He}{i} lines in the $I$-band are now almost vanished except for
the line at 8780\AA, still moderately strong.
In W21, we know that it does not show conspicuous \ion{Si}{iii}~5740,
confirming that it is earlier than B1. Another similar spectrum is
W61a, showing pure emission in H$\alpha$. \ion{C}{iii}, if present, is
very weak. We classify these three objects as B0.5\,Ia.

W43a shows pure emission in H$\alpha$. The Paschen lines are shallower
and we can only see them up to
Pa~18. \ion{C}{iii}~8500\AA\ is starting to contribute strongly to Pa~16. There is no sign of \ion{C}{ii} or \ion{N}{ii}
absorption now, but strong \ion{C}{ii} wind emission lines. This suggests a spectral type B0\,Ia. This object is a single-lined spectroscopic binary \citep{ritchie09a}. W34 has a similar
spectrum and receives the same classification.

W232 and W54 have broader Paschen lines, even though we see up to
Pa~18. W54 presents a neutralised H$\alpha$, while W232 has it weakly
in absorption. It also shows a very weak \ion{Si}{iii}~5740\AA\
line. The lack of H$\alpha$ emission likely indicates a lower mass loss and
hence luminosity. We  classify these objects as B0.5\,Iab. Another
similar spectrum is W6a, which shows variable H$\alpha$ and hard X-ray emission \citep{clark08}, and is a likely binary (see below). As discussed in Appendix~A, the strength of H$\alpha$ emission correlates well with luminosity, but there are many exceptions. 

W60 and W63a have similar spectra, with H$\alpha$ neutralised or weakly in absorption, but show clear evidence of \ion{C}{iii}~8500\AA. Paschen lines are broad, and we classify these two objects as B0\,Iab. Based on the overall similarity, we classify W49, for which we only have a FLAMES $I$-band spectrum, as B0\,Iab, as well.

W61b also has H$\alpha$ weakly in absorption, but has a prominent \ion{C}{iii} line, while Paschen lines are visible up to Pa~18. We classify this object as O9.5\,Iab, though a higher luminosity cannot be ruled out. Two other stars look like high-luminosity O-type supergiants. W74 has a very prominent \ion{C}{iii}~8500\AA\ line, and Paschen lines up to Pa~18. Its H$\alpha$ line is weakly in absorption. We classify it as O9.5\,Iab, though it also could be slightly earlier or slightly more luminous. W24 is very similar, with H$\alpha$ more clearly in absorption, and its Paschen lines show that it is either less luminous or earlier than the other two. This object shows some spectral variability \citep{clark09} and needs further study. We provisionally classify it as O9\,I.

Similar spectra, with H$\alpha$ close to being neutralised, are shown by W35 and W41. As they have stronger \ion{C}{iii}~8500\AA\ and Pa~18 is not clearly detected, we classify them as O9\,Iab. Based on overall similarity, we also classify as O9\,Iab a number of objects with lower-resolution (NTT) or incomplete (only $I$-band) spectra, W17, W25 and W38, though these classifications are less secure. W27 has a similar spectrum, but shows several indications of being a binary (complex H$\alpha$, shallow broad Paschen lines, likely double \ion{He}{i}~6678\AA), in accord with its high X-ray luminosity \citep{clark08}.

W30a displays strong variability and clear signs of binarity \citep[see also][]{clark08}.  Some of the spectra suggest that \ion{He}{ii}~8236\AA\ could be present, while \ion{C}{iii}~8500\AA\ changes between spectra. The Paschen lines are broad and shallow, typical of a spectroscopic binary. This system contains two O-type supergiants, and one of them could be earlier than O9.

For W31, we only have spectroscopy at one epoch, but the double-peaked structure of H$\alpha$ strongly suggests that this B0\,I supergiant has a companion, rendering the exact luminosity classification impossible. W10, as shown in \citet{clark08}, is a double-lined spectroscopic binary, and therefore must contain two supergiants.

To the South of the cluster, W238 does not seem to fit exactly with the spectral sequence.  It has \ion{C}{ii} in
absorption and H$\alpha$ partially neutralised, suggesting a spectral type B1\,Iab or B1\,Ib. However, $\lambda$8502\AA\ looks stronger than expected, and this could also be a composite spectrum..

All the other objects have H$\alpha$ in absorption and a smaller
number of Paschen lines. Based on the relative strength of the \ion{C}{iii}~8500\AA\ feature, we classify them O9-O9.5\,Ib. We consider that having H$\alpha$ in absorption to the point that it is stronger than the
\ion{He}{i}~6678\AA\ line is the defining characteristic of luminosity class Ib. However, we caution that sky subtraction is
difficult in the very crowded regions close to the cluster core, and H$\alpha$ could be affected by the sky.

Taking all these factors into account, we estimate that the spectral types we derive are likely accurate to within half a subtype for B supergiants. Luminosity classes are also likely to be correct to within one subclass. Types for O
stars are not so well determined, though we expect to be correct within one
subtype. Luminosity classes are less reliable, as we lack proper
criteria. We have to resort to line widths and shapes, which are also
dependent on rotational velocity. As the rotational velocities of O
supergiants may be relatively high, there is a degree of degeneracy
between luminosity and rotation. Also, the types derived depend
strongly on the ratio of the \ion{C}{iii}~8500\AA\ line to
neighbouring Paschen lines. This indicator is obviously sensitive to
CNO anomalies. However, we do not find any spectrum where the strength
of this line is openly in contradiction with any other criteria. Based
on this, we believe that any CNO anomalies present in our stars are,
at most, moderate. In the case of moderate C enhancement, a star
classified as O9\,Iab might actually be a B0\,Ib with enhanced
\ion{C}{iii}. In the case of moderate N enhancement (and, hence, C
depletion), a star classified B0-0.5\,Iab or Ib could be an O9\,Ia
with a very weak \ion{C}{iii}. None of these possible
misclassifications looks too severe.  

Finally, 
Table~1 lists the spectral types of 55 OB supergiants, and two less luminous objects, in Wd~1

\subsection{Bright B supergiants}

Two B5\,Ia and one B9\,Ia supergiants show characteristics that lead us to classify them as hypergiants. For a start, they are $>1$~mag brighter (after correcting for extinction; see below) than any other B-type supergiant. In addition, they show signs of heavy mass loss, such as strong P-Cygni profiles in H$\alpha$ (Fig.~\ref{fighyperalpha}) and the presence of a P-Cygni profile in \ion{O}{i}~8446\AA\ (Fig.~\ref{hyperblues}). \citet{clark09} present evidence for variability in these three objects. At this stage, we do not intend their classification as hypergiants to have any implication about their evolutionary status. Most of the B5--9\,I supergiants that have been subject to detailed analysis fall on tracks corresponding to stars with $\la25\:M_{\sun}$ \citep{mcerlean,markova}, and hence it seems natural to find that the late B supergiants in Wd~1, most likely descendants of more massive progenitors, are more luminous. Analysis of their chemical compositions will cast light on their evolutionary status.

\subsection{Transitional objects with emission lines}

\begin{figure}
   \centering
\resizebox{\columnwidth}{!}{\includegraphics[angle=-90]{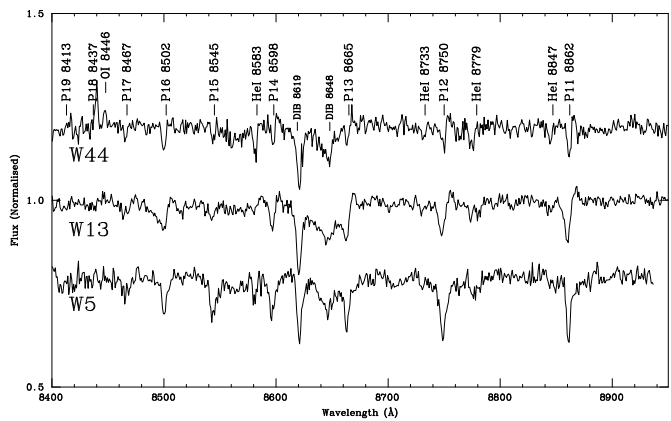}}
\caption{$I$-band spectra of three transitional objects with emission
  lines. W44 (=WR L) is a late Wolf-Rayet star, presenting emission components
  in all Paschen lines. W13 is an eclipsing binary. One of the
  components is the emission-line object. W5 shows no emission lines
  in the $I$-band in spite of strong emission features in the red (see Fig.~\ref{fighyperalpha}.}
              \label{transits}
    \end{figure}

Two objects displaying strong H$\alpha$ emission and \ion{He}{i} emission can be classified as early B objects, W5 and W13. Their H$\alpha$ profiles can be seen in Fig.~\ref{fighyperalpha} (wider range spectra are displayed in \citealt{clark08}). Their $I$-band spectra are displayed in Fig.~\ref{transits}, where they are compared to the WN9 star W44 (= WR L). 

W5 displays very intense H$\alpha$ emission, P-Cygni profiles in \ion{He}{i}~$\lambda\lambda$ 5785, 6678 \& 7065,  \ion{C}{ii} wind features (not only the 7231, 7236\AA\ doublet is in emission, but also 6578, 6583\AA) and some \ion{N}{ii} lines weakly in emission. In contrast, its $I$ band spectrum lacks any emission features and seems typical of a supergiant around B0.5 (though weak emission components in the Paschen lines may not be seen at our resolution). In view of this, we classify this object as a B0.5\,Ia$^{+}$ star, where the hypergiant classification is due to the several indicators of heavy mass loss. However, we note that this object is intrinsically {\em less} luminous than other stars classified as B0.5\,Ia. In view of this and its similarities to WN9, it is tempting to identify W5 with an evolutionary phase immediately prior to the WR stage. Again, a careful analysis of its chemical composition may provide further clues as to its evolutionary status.

W13 has a similar spectrum, but analysis is at present complicated by the fact that this object is a double-lined eclipsing binary \citep{bonanos,ritchie09a}. Pending a detailed analysis of the system, the data seems consistent with an emission-line component similar to W5 and an absorption-line object. Though the Pa~16 feature in Fig.~5 is very prominent, several FLAMES spectra \citep[cf.][]{ritchie09a} show moderately strong \ion{He}{i} lines, indicating that the companion is likely an early-B supergiant.

  \begin{table*}
\label{longlist}
\begin{center}
\caption{Sample of OB supergiants in Wd 1, with $BVRI$ photometry from
  \citetalias{main} (when available) and spectral types\fnmsep$^{a}$.}
\begin{tabular}{lccccccc}
\hline
ID    & RA (J2000) & Dec (J2000) & $B$ & $V$ & $R$ & $I$ &  Spectral \\ 
     &  &            &             &   &   &      &  Type\\     
\hline
W2a   &  16 46 59.71 & $-$45 50 51.1 & 20.4 & 16.69 & 14.23 & 11.73 & B2\,Ia\\ 
W5   &  16 47 02.97 & $-$45 50 19.5 & 21.4 & 17.49 & 14.98 & 12.48 & B0.5\,Ia$^{+}$\\
W6a   &  16 47 03.04 & $-$45 50 23.6 & 22.2  & 18.41 & 15.80 & 13.16 & B0.5\,Iab\\
W6b$^{**}$ &16 47 02.93 & $-$45 50 22.3 &23.6 &20.20&17.91&15.25&O9.5\,III\\
W7   &  16 47 03.62 & $-$45 50 14.2 & 20.0 & 15.57 & 12.73 & 9.99 & B5\,Ia$^{(+)}$\\
W8b  &16 47 04.95 &$-$45 50 26.7 &$-$ &$-$ &$-$ &$-$ & B1.5\,Ia \\
W10  &  16 47 03.32 & $-$45 50 34.7 &  $-$    &   $-$   &   $-$   &   $-$   & B0.5\,I+OB\\ 
W11  &  16 47 02.23 & $-$45 50 47.0 & 21.2 & 17.15 & 14.52 & 11.91 & B2\,Ia   \\
W13  &  16 47 06.45 & $-$45 50 26.0 & 21.1 & 17.19 & 14.63 & 12.06 & B0.5\,Ia$^{+}$+OB\\
W15  &  16 47 06.63 & $-$45 50 29.7 & 22.8  & 18.96 & 16.38 & 13.75 &  O9\,Ib   \\
W17$^{**}$  & 16 47 06.25& $-$45 50 49.2 & 22.7 & 18.87& 16.19 & 13.56 & O9\,Iab   \\
W18& 16 47 05.71 & $-$45 50 50.5&21.2&17.32&14.81&12.27&B0.5\,Ia\\
W19  &  16 47 04.86 & $-$45 50 59.1 & 22.6  & 18.22 & 15.21 & 12.37 & B1\,Ia\\ 
W21 &  16 47 01.10& $-$45 51 13.6&22.5&18.41&15.56&12.74&B0.5\,Ia\\
W23a  &  16 47 02.57 & $-$45 51 08.7 & 22.1  & 17.85 & 14.91 & 12.07 & B2\,Ia+B\,I?\\
W24  &  16 47 02.15 & $-$45 51 12.4 & 23.0  & 18.71 & 15.96 & 13.24 & O9\,Iab\\
W25$^{*}$& 16 47 05.78& $-$45 50 33.3&21.9&17.85 &15.22  & 12.61& O9\,Iab \\
W27$^{*}$& 16 47 05.15& $-$45 50 41.3&21.5&17.94&15.35&12.80&O9\,I+O\,I \\
W28  &  16 47 04.66 & $-$45 50 38.4 & 20.9 & 16.87 & 14.26 & 11.64 & B2\,Ia \\
W29  &  16 47 04.41 & $-$45 50 39.8 & 22.6  & 18.66 & 16.02 & 13.38 & O9\,Ib\\
W30  &  16 47 04.11 & $-$45 50 39.0 & 22.4  & 18.45 & 15.80 & 13.20 & O+O\\
W31 & 16 47 03.78 &$-$45 50 40.4&$-$&$-$&$-$&$-$& B0\,I+OB\\
W33  &  16 47 04.12 & $-$45 50 48.3 & 20.0 & 15.61 & 12.78 & 10.04 &  B5\,Ia$^{(+)}$\\
W34 &16 47 04.39 &$-$45 50 47.2&22.1&18.15&15.40&12.69&B0\,Ia\\
W35 &  16 47 04.20 & $-$45 50 53.5 & 22.7 & 18.59 & 16.00 & 13.31 & O9\,Iab\\
W37$^{**}$ &16 47 06.01&$-$45 50 47.4&22.8&19.11&16.40&13.65&O9\,Ib\\
W38$^{**}$ &16 47 02.86&$-$45 50 46.0&23.2&19.10&16.47&13.81&O9\,Iab\\
W41  & 16 47 02.70  & $-$45 50 56.9 & 21.3  & 17.87 & 15.39 & 12.78 & O9\,Iab\\
W42a  &  16 47 03.25 & $-$45 50 52.1 &  $-$    &   $-$   &   $-$   &   $-$   & B9\,Ia$^{(+)}$ \\
W43a  & 16 47 03.54  & $-$45 50 57.3 & 22.8  & 18.05 & 15.22 & 12.26 & B0\,Ia\\ 
W43b  & 16 47 03.52 & $-$45 50 56.5&    $-$    &   $-$   &   $-$   &   $-$   &B1\,Ia\\ 
W43c & 16 47 03.76 & $-$45 50 58.3& 20.4& 18.35  &16.18  & 13.66 & O9\,Ib\\ 
W46a& 16 47 03.91& $-$45 51 19.5&23.0&18.55&15.46&12.46& B1\,Ia\\
W46b& 16 47 03.61& $-$45 51 20.0&$-$&$-$&$-$&$-$&O9.5\,Ib\\
W49$^{**}$ &16 47 01.90 &$-$45 50 31.5&22.6&18.76&16.30&13.80&B0\,Iab\\
W50b$^{**}$ &16 47 01.17&$-$45 50 26.7 &22.8 &19.66&17.21&14.69&O9\,III\\
W52 &16 47 01.84 & $-$45 51 29.2&21.8&17.48&14.68&11.94&B1.5\,Ia\\
W54&16 47 03.06 & $-$45 51 30.5&$-$&$-$&$-$&$-$&B0.5\,Iab\\
W55$^{**}$ &16 46 58.40&$-$45 51 31.2 &21.6&17.67&15.25&12.67&B0\,Ia\\
W56b$^{**}$ &16 46 58.85&$-$45 51 45.8&22.8&18.88&16.36&13.76&O9.5\,Ib\\
W57a  & 16 47 01.35  & $-$45 51 45.6 & 20.7 & 16.54 & 13.83 & 11.13 & B4\,Ia\\
W60  & 16 47 04.13  & $-$45 51 52.1 & 22.8  & 18.50 & 15.96 & 13.28 & B0\,Iab\\
W61a  & 16 47 02.29  & $-$45 51 41.6 & 21.2 & 17.16 & 14.62 & 12.01 & B0.5\,Ia \\
W61b  & 16 47 02.56  & $-$45 51 41.6 & 22.7 & 18.59 & 16.00 & 13.31 & O9.5\,Iab\\
W62a  & 16 47 02.51  & $-$45 51 37.9&    $-$    &   $-$   &   $-$   &   $-$   &B0.5\,Ib\\
W63a &16 47 03.39&  $-$45 51 57.7 &22.6   &18.56 & 16.20  & 13.68  &B0\,Iab\\
W65$^{**}$ &16 47 03.89&$-$45 51 46.3&22.9&18.73&16.27&13.68&O9\,Ib\\
W70  & 16 47 09.36  & $-$45 50 49.6 & 21.2 & 16.88 & 14.10 & 11.29 & B3\,Ia \\
W71 & 16 47 08.44  & $-$45 50 49.3 & 21.5 & 17.01 & 14.06 & 11.16 & B2.5\,Ia \\
W74 &16 47 07.08  &$-$45 50 13.1  &    $-$    &   $-$   &   $-$   &   $-$   &O9.5\,Iab\\
W78 &16 47 01.54 &$-$45 49 57.8 &21.0   & 17.06  & 14.54 &12.04  &B1\,Ia\\
W84$^{**}$ &16 46 59.03&$-$45 50 28.2&21.3&17.82&15.60&13.63&O9.5\,Ib\\
W86$^{**}$ &16 46 57.15&$-$45 50 09.9&22.9&18.76&16.43&14.00&O9.5\,Ib\\
W228b & 16 46 58.05 & $-$45 53 01.0 &$-$    &   $-$   &   $-$   &   $-$& O9\,Ib\\
W232 &  16 47 01.41 & $-$45 52 34.9 &21.3  &17.53  &15.25 &12.85 & B0\,Iab\\
W238 &  16 47 04.41 & $-$45 52 27.6 & 21.4 & 17.47 & 14.98 & 12.45 & B1\,Iab\\
W373$^{**}$ &16 46 57.71 & $-$45 53 20.1& $-$ &$-$&$-$&$-$ &   B0\,Iab     \\
\hline
\end{tabular}
	\begin{list}{}{}
\item[]$^{a}$ Objects marked with ``$^{*}$'' have classifications based on EMMI low-resolution (grism \#6) spectra. Their spectral types are therefore less accurate.\\
Objects marked with ``$^{**}$'' have classifications based only on FLAMES spectra covering 8480--8920\AA. Their spectral types are tentative.
\end{list}
\end{center}
\end{table*}

%
%

\subsection{Likely binaries}

Several spectra show evidence for binarity. In W10 we see double
\ion{He}{i} lines. In W23a, we see features incompatible with a single
spectral type and variations between two epochs. W30a and W31 show
evidence of mass transfer in the form of complex and (in W30a) variable H$\alpha$. W30a and W27 show shallow lines, typical of double-lined binaries. W6a also shows important variations in the shape of H$\alpha$
\citep{clark08} and its FLAMES spectra show very shallow broad lines. The spectra of W27 have lower resolution than
those of other sources, but the shape of the \ion{He}{i} lines is also
suggestive of two components, while H$\alpha$ shows a complex emission
profile. Of these, W6a, W10, W27 and W30a are hard X-ray sources
\citep{clark08}, the latter being the most luminous and hardest
detection in the cluster after the magnetar CXO~J164710.2$-$455216
\citep{clark08}. W27 is also a relatively bright X-ray source
($L_{{\rm X}}\sim10^{33}\:{\rm erg}\,{\rm s}^{-1}$).

 Considering the relative brightnesses of OB supergiants and MS stars, every object showing double lines must be considered to contain two supergiants. Given the relative shortness of this evolutionary phase, the presence of at least two (and likely more)
objects displaying the spectra of two supergiants suggests that there
is a high number of systems which actually are similar-mass
binaries, in good agreement with the preliminary results of a radial velocity survey, designed to determine the binary fraction in Wd~1 \citep{ritchie09a}.

\subsection{The less luminous objects}

\begin{figure}
\centering \resizebox{\columnwidth}{!}{\includegraphics[angle=-90]{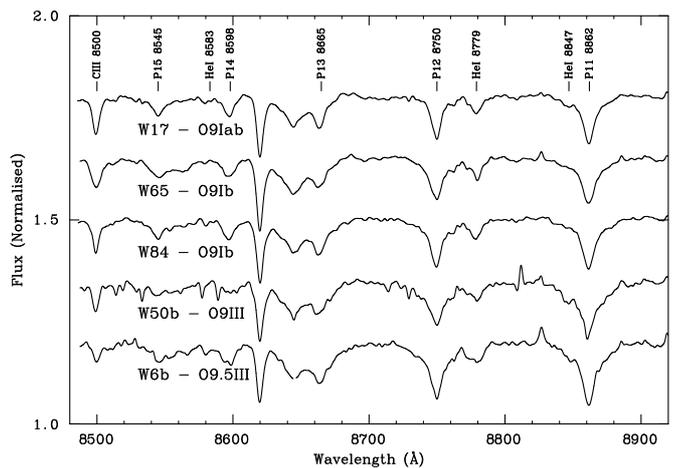}} 
 \caption{Sequence of FLAMES/GIRAFFE $I$-band spectra of some of the earlier spectral types observed in our sample.  The bottom two spectra have lower SNR, but the Paschen lines are clearly seen to have much broader wings than the more luminous objects and display much weaker \ion{C}{iii}~8500\AA. Note the difficulty in defining the continuum around the area of the DIBs at 8620\AA\ and 8648\AA, in spite of the high resolution. The spikes seen in some spectra are badly removed sky lines.}
   \label{giants}
\end{figure}

In order to illustrate the characteristics of O-type stars in the $I$-band, we have included in our sample two stars with significantly fainter $I$ magnitude. These are W6b and W50b. Their spectra are displayed in Fig.~\ref{giants}, compared to some other objects that were observed only with FLAMES. The small spectral coverage makes all classifications somewhat uncertain, but Fig.~\ref{giants} shows that a few spectral features may still provide some guidance. The main discriminant for luminosity class is the shape and depth of the Paschen lines.

The two faint stars have significantly broader Paschen lines than any star that we classify as supergiant. The limit between the wings of the lines and the continuum is difficult to define, in opposition to the sharply defined lines of the supergiants. Lacking any further criteria, we tentatively classify these two stars as luminosity class III, noting that they may well be of lower luminosity -- as suggested by their dereddened magnitudes (see below). They are representative of the large population of unevolved massive stars present in Wd~1, which we will study, with a more extensive dataset, in a future paper.

\section{Discussion}

We have spectroscopically identified 55 OB supergiants in
Wd~1. The sample is still very far from complete. On the one hand, the
extended halo around the cluster is not fully explored. On the other
hand, many stars in the cluster core, with magnitudes and colours
similar to the O-type supergiants listed here, still lack spectroscopy. 

The significance of this huge population has to be discussed within the context of their membership in a single cluster, Wd~1, which hosts a large population of massive stars in more advanced evolutionary stages. The cluster contains one probable active hot LBV \citep[W9;][]{dougherty}, 7 A/F hypergiants (including the cool LBV W243; \citealt{ritchie09b}), 4 very luminous RSGs \citep{clark09} and a large population of WR stars \citep{crowther}. The OB supergiants provide us with the possibility to constrain the evolutionary paths of these evolved stars. Unfortunately, the parameters of Wd~1 are still poorly determined, even if significant progress has been made in recent years. The data presented here can set strong constraints on many of these parameters.

\subsection{Extinction}

The distance to the cluster is, at present, not very well constrained. Photometry is affected by very strong reddening. Analysis of the $E(B-V)$ colours
suggested that the 
reddening could deviate from the standard law \citepalias{main}, with an extinction $A_{V}\approx12\:{\rm mag}$. With the new accurate spectral types, we can conduct a deeper investigation of the reddening to the cluster.

\begin{figure}
\centering
 \resizebox{\columnwidth}{!}{\includegraphics[angle=0, clip]{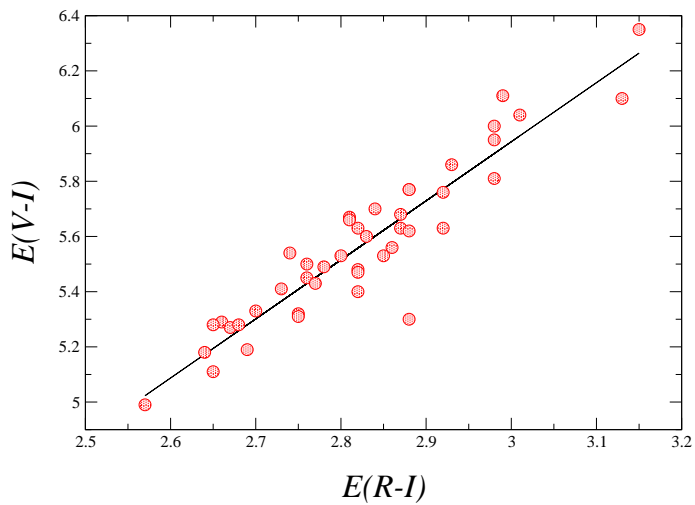}} 
 \caption{Correlation between the $E(V-I)$ and $E(R-I)$ colour excesses for OB supergiants in Wd~1. Individual excesses have been derived from the photometry listed in Table~1 and the intrinsic colour calibration.}
   \label{fig:excesses}
\end{figure}

We have taken all the objects with photometry in Table~1 and calculated their colour excesses, making use of the intrinsic colours for OB supergiants \citep{wegner}. Figure~\ref{fig:excesses} shows the run of $E(R-I)$ against $E(V-I)$. The two colours cover a wide range, showing that the reddening is very variable across the face of the cluster. $E(V-I)$ goes from $\sim5.0$~mag to $\sim6.4$~mag, with a single object, W84, showing rather lower reddening $E(V-I)=4.5$. The two colours are very well correlated ($R=0.99$ for a fit without independent term). The best fit gives $E(V-I)/E(R-I)=1.97\pm0.01$, which compares very well with the standard reddening law \citep[e.g., 1.95 in][]{rieke}. The average value is $E(V-I)=5.5\pm0.3$ (1-$\sigma$ standard deviation). Similarly good fits are obtained for $E(V-I)/E(V-R)$ and $E(R-I)/E(V-R)$ (which, obviously, are not independent), giving values very close to the standard reddening law.

This suggests that the extinction is close to standard, in spite of the difficulty with the $E(B-V)$ colours. The correlation between $E(B-V)$ and $E(V-I)$ is poorer.  If we force the fit to go through the origin, we obtain $E(V-I)/E(B-V)=1.31\pm0.01$, with $R=0.97$. This value is almost identical to the average of the ratios $E(V-I)/E(B-V)$ for all the stars, $1.34\pm0.15$. The value is not compatible with the standard value, \citep[e.g., 1.6 in][]{rieke}, but, because of the large dispersion in individual values, is less than $2\sigma$ away. In view of this, we suspect that the difficulty stems from unreliable values for $B$, perhaps a zero-point offset. Even though the photometry in \citetalias{main} was very carefully calibrated, stars in Wd~1 are much redder than any standards used, and have very faint $B$ magnitudes. In any event, the average value directly determined from the observed $(B-V)$ colours is $E(B-V)=4.2\pm0.4$ (again, 1-$\sigma$ standard deviation).

Another reliable indicator of interstellar extinction is the presence of Diffuse Interstellar Bands (DIBs). Unfortunately, the correlations between the Equivalent Width (EW) of most bands and reddening saturates at relatively low values \citep[e.g.,][]{mz97,cox05}. \citet{munari08} find that the 8620\AA\ DIB is an excellent tracer of reddening, with a very tight correlation $E(B-V) = 2.72(\pm0.03) \times~$EW~(\AA). As discussed in the appendix, measurement of the intensity of this band is not easy at our resolution. However, we may expect the high number of objects to compensate for the uncertainty in individual values. Unfortunately, there is no obvious correlation between EW(DIB) and $E(V-I)$ for our stars. Objects in Wd~1 show an average value EW(DIB)$=1.00$\AA, with a dispersion $\sigma=0.09$ and individual values ranging from 0.85 to 1.15, with no correlation to reddening or spatial position. 

If we restrict the search to objects with high-resolution FLAMES spectra, we find a similar result. The average is now $1.11$\AA, showing that the lower resolution data tend to give lower values because of the poorer definition of the continuum. The dispersion is only $\sigma=0.08$, but still no correlation with $E(V-I)$ or location is obvious.

This suggests that this DIB is saturated, at least along this line of sight, at the reddening of Wd~1, and the dispersion amongst Wd~1 stars is due simply to the difficulty in determining the continuum. As discussed in the appendix, this area contains a broad shallow DIB centred on 8648\AA\ \citep[cf.][]{munari08}, which becomes compounded with some weak absorption lines. The correlation of \citet{munari08} would give an average $E(B-V)=3.00$, which is smaller than both the observed value and a direct extrapolation from $E(V-R)$ and $E(V-I)$ under the assumption of a standard reddening law. 

There is another sharp DIB in our range, that at 6614\AA. Unfortunately, its correlation to $E(B-V)$ seems to be poor at high reddening values \citep{cox05}. Again, we find no obvious correlation to $E(V-I)$. Most stars show EW(DIB)$=0.60$\AA, with standard deviation $\sigma=0.05$. As this value is comparable to the highest EW(DIB) measured by \citet{cox05} along sightlines with high extinction, it is possible that this line is also saturated, though measurements at higher resolution should be desirable to check this hypothesis.

The only DIB that has been calibrated against $E(B-V)$ and might perhaps not be saturated at the reddening of Wd~1 is the narrow band at 5797\AA. Unfortunately, this is just outside the range of our spectra. It is seen in a few of the MXU spectra, due to their variable spectral range, but the continuum cannot be defined well.

\subsection{Distribution of spectral types}
The earliest supergiants in Wd~1 seem to have spectral types around O9. This is
defined by the presence of strong \ion{C}{iii}~8500\AA\ (while the
neighbouring Paschen lines are weak or absent) and lack of
\ion{He}{ii}~8238\AA. Even though our spectra have not been corrected
for telluric absorption, humidity was very low on the night of June
13th, and the stronger \ion{He}{ii} line of earlier spectral types should be
noticeable. None of the stars classified as O9 or O9.5 seems
sufficiently luminous to receive a Ia luminosity class, based on their
relatively broad Paschen lines and lack of 
detection beyond Pa~17 (but see below). W74 (where Pa~18 is weakly seen), is the most luminous O-type supergiant based on these criteria.  

There are a fair number of luminous (Ia) supergiants covering the range
B0.5--B2. Later spectral types become less numerous, with one each of
B2.5\,Ia, B3\,Ia and B4\,Ia, and two B5\,Ia$^{(+)}$ bright
supergiants. After this, we only have a B9\,Ia$^{(+)}$ object before we reach the A hypergiants (which include the present phase of the LBV W243; \citealt{ritchie09b}). This distribution hints at some sort of break around spectral type B2--3, perhaps suggesting
that stars in this mass range finish their H-core burning phase as
$\sim$B2 supergiants.

Though our sample of OB supergiants is far from complete, there have been no systematics in the choice of targets. The level of incompleteness is difficult to assess, as photometry is not available for all the stars in the crowded cluster core area. We estimate that perhaps $\sim10$ stars which appear bright enough in $I$ to be supergiants have not been observed in the cluster core (most notably W12b, W39a \& W40a), with a similar number in the outskirts (e.g., W53 or W56a). Of course, a much larger population of less luminous OB star remain to be explored. 

\subsection{Mass}

The mass of Wd~1 has been estimated in two different ways. \cite{mt07} have used the radial velocity dispersion of the ten stars brightest in the infrared
($\sigma=8.4\:{\rm km}\,{\rm s}^{-1}$) to estimate a
mass of  $\sim 1.3\times10^{5}\:M_{\odot}$. This estimate needs to be
confirmed, as it is subject to a large uncertainty. The YHGs show
large variations in their radial velocities due to pulsations
\citep{ritchie09a, clark09}, and some of the stars used present important
spectral peculiarities (e.g., W26 is associated with extended nebular
emission). Because of this, and the assumption of virial equilibrium, this value should be taken as an upper limit. \citet{brandner}, on the other hand,  have used star counts in the infrared to set a lower limit on the cluster mass  $M_{\rm  cl}\ga5\times10^{4}\:M_{\odot}$. Again, 
there are important uncertainties coming into this estimation, such as the completeness correction.  

In particular, the pre-main-sequence isochrones used by \citet{brandner} indicate a significantly younger ($\tau=3.3\pm0.2$~Myr) and less distant ($d=3.6\pm0.2$~kpc) cluster. This results in very different masses for a given star. Recently, \citet{naylor} has shown that similar discrepancies are observed for a large number of young open clusters and suggested that this effect reflects a real deficiency in the calculation of pre-main-sequence ages. 

In any case, the integrated initial mass
function, which is independent of the distance and age estimate, appears consistent with Salpeter's down to solar mass \citep{brandner}. The spectroscopic confirmation of a large number of photometric candidate members from \citetalias{main} here indicates that the number of stars more massive than the main sequence turn-off is likely to be $\sim 200$. If we accept that the turn-off stars have $M_{*}\ga25\:M_{\sun}$, a Kroupa IMF would then imply a mass  $\sim 10^{5}\:M_{\odot}$, consistent with both determinations.

An important source of uncertainty in both measurements is the role of
binarity, which affects both radial velocities and transformation of
magnitudes into masses. Different lines of evidence support a very high fraction of binaries with similar mass components amongst cluster
members. Most WR stars show indirect indications of binarity:
presence of dust in WC stars \citep{crowther} and hard X-ray emission
in WN stars \citep{clark08}. Some of them also show photometric
variability and at least one is an eclipsing binary \citep{bonanos}. The pilot study by \citet{ritchie09a} shows that the high binary fraction extends to lower masses.

\subsection{Distance and age}

A recent
determination of the distance to Wd~1, making use of atomic hydrogen in the direction to the cluster, gives $d=3.9\pm0.7$~kpc \citep{kothes}, compatible with, though slightly shorter than, estimates based on the stellar population
\citep[e.g.,][]{crowther}. If we use the standard reddening law to deredden the magnitudes and assume $DM=13.0$, we find that luminosity class Ia stars typically have $M_{V}$ between $-6.4$ and $-6.9$\fnmsep\footnote{Though a few are much fainter. In particular, W55 has $-5.6$, introducing some doubts about its classification, which is based only on a FLAMES spectrum.}. This is not in bad agreement with existing calibrations \citep[e.g.,][]{hme84}, though the slightly longer distance of \citet{crowther} ($DM\approx13.4$) would give a better agreement. The three mid-B supergiants W57a, W70 and W71 are definitively brighter, with $-8<M_{V}<-7$. W7 and W33 are even brighter, with $M_{V}\approx-8.5$ (which would become almost $-9$ for $DM=13.4$), justifying their tentative classification as hypergiants.

\begin{figure}
\centering \resizebox{\columnwidth}{!}{\includegraphics[angle=0,clip]{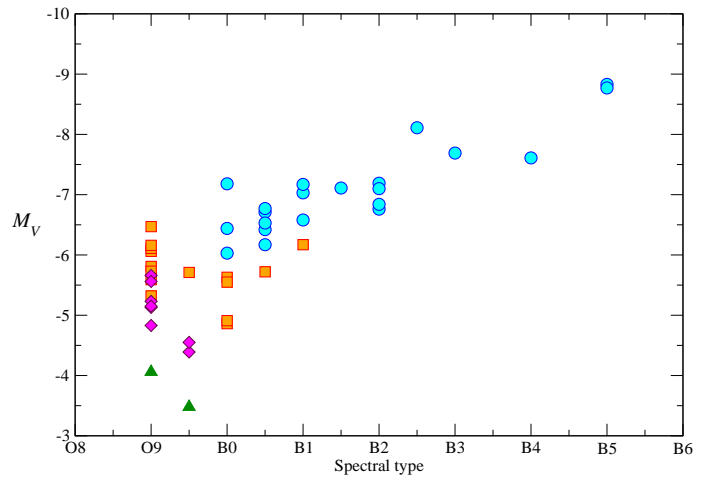}} 
 \caption{Observational HR diagram for Wd~1. The absolute magnitudes, $M_{V}$, are estimated from the observed $I$, dereddened using $E(V-I)$ under the assumption of a standard law and corrected with the tabulated $(V-I)_{0}$ for the spectral type from \citet{wegner}. Circles represent objects classified as Ia supergiants; squares are Iab classifications, diamonds are Ib objects and triangles are stars with lower luminosity.}
   \label{fighr}
\end{figure}

In Fig.~\ref{fighr}, we plot an estimate of the intrinsic luminosity for all the objects with good photometry in Table~1 distributed by spectral type and luminosity class. For this, we use $E(V-I)$, under the assumption of a standard reddening law \citep{rieke}, to calculate $A_{I}$, and then derive $M_{I}$ for a nominal $DM=13.5$. Due to the uncertainty in $E(B-V)$, we prefer to perform our calculations with the red magnitudes. However, for ease of comparison, we transform to $M_{V}$ simply by adding the intrinsic $(V-I)_{0}$ tabulated for the spectral type. The plot shows, in general, a good correlation between absolute magnitude and luminosity class, giving strong support to our classifications. The obvious exception are the objects classified as O9 SGs. These objects cover a broad span ($>2$~mag) in $M_{V}$, present an overlap between O9\,Iab and O9\,Ib objects (though objects classified O9\,Iab are in most cases brighter than objects classified O9\,Ib) and have $M_{V}$'s rather higher than expected (the O9\,Iab objects are as bright as the B0\,Ia stars).

This discrepancy may arise from the difficulty in separating the effects of luminosity and fast rotation in O-type stars at this moderately low resolution. Some of the objects classified as O9\,Iab may be fast-rotating O9\,Ia stars. Though plausible, this interpretation presents two difficulties: first, all O9\,Ia stars present should be fast rotators (perhaps not a serious objection when we are talking of 3--4 stars); second, it would mean that we have a very large range of luminosity classes at spectral type O9 (from O9\,III to O9\,Ia), as opposed to other spectral subtypes. Alternatively, as discussed above, we can think that some of the objects classified as O9\,Iab are really later-type ($\sim$B0) stars with enhanced \ion{C}{iii}~8500\AA, but this does not necessarily imply brighter $M_{V}$. Perhaps the simplest interpretation is that the objects we have classified as O9\,Iab are really O9.5\,Ia. As seen in Fig.~8, this small change in their classification would make their luminosities compatible with all the other members\fnmsep\footnote{As discussed in Section~3, many of the O9\,Iab classifications are less reliable than the rest, due to lower resolution or smaller spectral range observed.}.

With this caveat, we may try to constrain the age of the cluster from the observed population. For this, we are forced to rely on the predictions of evolutionary models. As discussed above, the tracks followed by high-mass stars in the theoretical HR diagram are subject to large uncertainties. As the exact mechanism for heavy mass loss is unknown, mass loss rates are normally introduced by hand, following some recipe. Differences in mass loss rates, the treatment of rotation and the degree of convective overshooting assumed lead to rather different evolutionary tracks \citep[e.g.,][]{sal99,mm,eldridge08}.  Furthermore, the inclusion of binarity can substantially alter the evolution of massive stars \citep[e.g.,][]{vanbeveren07,eldridge08}.

As discussed by \citet{crowther}, single evolutionary models have trouble reproducing the observed population of Wd~1. Geneva isochrones without rotation \citep{mm} predict that RSGs will be formed only for ages $\ga6$~Myr, while even higher ages are required to produce RSGs with the high-rotation isochrones. On the other hand, the WR populations decline rapidly after 5~Myr. The simultaneous presence of large numbers of cool super/hypergiants and WR stars is reproduced better by models including binaries, which also provide a better description of the ratio between WC and WN stars. Comparison of the number ratio between WR stars and cool hypergiants with isochrones from \citet{eldridge08} led \citet{crowther} to propose an age of 4.5 or 5.0~Myr, with the progenitors of
the WR stars having initial masses in the
40\,--\,$55\:M_{\odot}$ range.

Such age seems consistent with the observed distribution of blue supergiants. We observe O9--B0 supergiants with different luminosity subclasses, while at later spectral types we only see Ia supergiants. At $\sim4.5$~Myr, these Ia supergiants should be descended from stars with initial masses $M_{{\rm ini}}\sim 35\:M_{\odot}$. This is in good agreement with placement on theoretical tracks of B\,Ia supergiants. For instance, \citet{crowtherbsgs} find that a sample of B0--3\,Ia Galactic supergiants lie between the tracks for 25 and $40\:M_{\odot}$. 

As single star isochrones have been frequently used to date open clusters, here we will compare the properties of OB supergiants to Geneva isochrones. As the stars are still relatively unevolved, the differences with respect to other single star tracks are minor, allowing easy comparison to the ages of other open clusters.

\begin{figure}
\centering \resizebox{\columnwidth}{!}{\includegraphics[angle=0,clip]{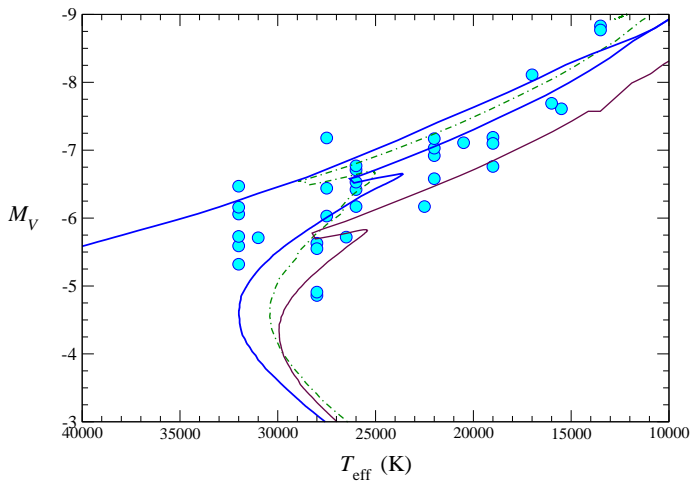}} 
 \caption{Semi-empirical HR diagram for bright OB supergiants in Wd~1. The solid lines represent the Geneva isochrones without rotation for $\log t =6.7$ (5~Myr; top, blue) and $\log t = 6.8$ (6.3~Myr; bottom, brown). The dash-dotted line is the $\log t = 6.8$ isochrone for high initial rotational velocity.  }
   \label{figtemps}
\end{figure}

In Fig.~\ref{figtemps}, we plot the same estimates of $M_{V}$ derived above against $T_{{\rm eff}}$ for all the Ia and Iab supergiants. The effective temperatures used are an interpolation between the observational scale of \citet{martins} for late O-type stars and the temperature scale of \citet{crowtherbsgs} for B-type supergiants. We also plot several isochrones from the models of \citet{mm}. Taking into account that the $M_{V}$ determination is subject to important uncertainties (for example, simply the rounding up of the second decimal in the determination of the ratio between $A_{I}$ and the colours amounts to a difference $>0.1$~mag) and the approximate character of the $T_{{\rm eff}}$ scale, the general agreement is rather good. We note the following points:
\begin{itemize}
\item The $\log t=6.7$ (5~Myr) isochrone without rotations would provide a good fit for a distance approaching 6~kpc.
\item The $\log t=6.8$ (6.3~Myr) isochrones (with and without rotation) provide a very good fit to the position of the B supergiants for distances $\sim4.5$--5~kpc. These isochrones would, however, fail to reproduce the position of the O supergiants and almost certainly fail to reproduce the turn-off of the main sequence, unless all the objects we have classified as O-type supergiants are blue stragglers.
\item The location of the mid-B supergiants is compatible with their being evolving away from the main sequence.
\item The distance from \citet{kothes} ($DM=13.0$) implies an age $>7$~Myr for the cluster, which seems incompatible with the population of WR stars. Their error bars, however, reach within the area favoured by our data.
\end{itemize}

If we accept that a significant fraction of our stars are binaries containing two massive stars, they would be intrinsically brighter, allowing a longer distance (by $\sim 0.5$~mag) for a given age.

Considering the uncertainties involved here, a definite answer will have to wait for accurate spectral classification of objects near the turn-off of the main sequence, which can be achieved with high quality infrared spectra. This will allow the anchoring of the infrared photometric sequence, and a much more accurate isochrone fit. The richness of the OB population in Wd~1 may then allow the creation of a more accurate temperature scale based on fits of tailored synthetic spectra to high quality spectra of our sample. Meanwhile, our data strongly favour $d\ga5$~kpc and $\tau\ga5$~Myr for Wd~1, and seem to definitely rule out the values of \citet{brandner}, based on pre-main-sequence tracks. As this paper contains a careful analysis of the data, we suspect that 
this result reflects inaccuracies in the pre-MS isochrones, as found by \citet{naylor} for other young open clusters.

\subsection{The yellow hypergiants}

Wd~1 contains six luminous hypergiants with spectral types between A5\,Ia$^+$ and F8\,Ia$^+$ \citep{clark09}. They represent a sizable fraction of the known Galactic YHGs. The evolutionary status of YHGs is unclear under current models. In several schemes \citep[e.g.,][]{smith04}, they are assumed to be stars that have already been RSGs and are coming back towards the blue in their way to become Wolf-Rayet stars. With the current dataset, the evolutionary status of the YHGs in Wd~1 is difficult to assess. A simple estimation of their luminosities using the same procedure utilised for the OB supergiants suggests that their bolometric luminosities are not very different from those of the B hypergiants (and, in the case of W8a, substantially lower)\fnmsep\footnote{Note that we lack photometry for such interesting objects as W42a (B9\,Ia$^{+}$), W32 (F5\,Ia$^{+}$) and W243 (LBV/A3\,Ia$^{+}$).}.

 We have to be careful with this estimation, though. Spectroscopic monitoring of the YHG W265 has shown it to present strong pulsational variability, which leads to changes in spectral type between F1\,Ia$^{+}$ and F5\,Ia$^{+}$, with a period of $\sim100$~d \citep{clark09}. There is a high likelihood that the other YHGs experiment similar variations. Such variability in spectral type should be reflected in significant changes in the intrinsic colours and bolometric corrections. Our photometric and spectroscopic data are not simultaneous, and so the two datasets may correspond to different spectral types. Moreover, and perhaps more importantly, strong pulsations may lead to heavy mass loss, and the assumption of typical colours for A--F supergiants may not be appropriate if the stars suffer from intrinsic absorption. 

Therefore a detailed spectroscopic analysis, including model fits for abundance determination, seems instrumental in deriving the evolutionary status of the YHGs. So far, such an analysis has only been conducted for the LBV/A3\,Ia$^{+}$ W243, which shows very clear signs of heavy element processing in its atmosphere. As noted by \citet{crowther}, single star evolutionary models do not reproduce the simultaneous presence of luminous cool supergiants and WR stars in Wd~1. Models including binary evolution fare better in this respect but, as pointed out by \citet{clark09}, no current model seems able to accommodate the large population of cool supergiants at the estimated age. Understanding the evolutionary relations between the evolved stars in Wd~1 will therefore provide very strong constraints on evolutionary models. 

\subsection{Spatial distribution}

Based on star counts in the infrared, \citet{brandner} found the cluster to be clearly elongated. The diffuse X-ray emission from the cluster shows a similar elongation \citep{muno06}, with the major axis at Position Angle $13\degr\pm3\degr$. 

\begin{figure*}
\centering \resizebox{\textwidth}{!}{\includegraphics[angle=90,clip]{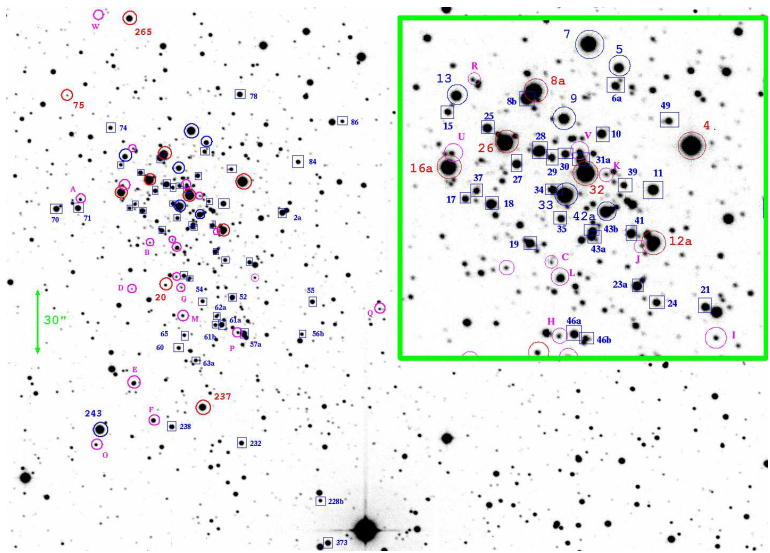}} 
 \caption{$R$-band finder ($\sim6\arcmin\times4\arcmin$) for all the stars classified as OB supergiants in this paper. Other stars in more advanced evolutionary stages are also shown. The moderately evolved OB supergiants are shown as rectangles. The more evolved stars are shown as circles: blue circles are blue hypergiants and LBVs, red circles are A/F hypergiants and M supergiants, while magenta circles are WR stars. The inset ($\sim70\arcsec\times70\arcsec$) shows the finder of the crowded cluster core. Many stars in this area (and also in the halo) have not been observed. Note that WR stars N, X and T are outside the area covered by this finder.}
   \label{figfinder}
\end{figure*}

In Fig.~\ref{figfinder}, we compare the distribution of the OB SG to that of the more evolved stars. For this latter set (comprising A/F hypergiants, M supergiants and WR stars), we can be certain of having a complete sample. The distributions of the two sets are not significantly different, revealing 
a very compact core, which contains more than half the objects of interest. This core measures only $\sim60\arcsec\times45\arcsec$, with the centre approximately defined by the positions of W27, W29 and W30, and is elongated along the approximate NE-SW line. In addition, there is a secondary clump to the South, with a number of objects likely defining a more symmetrical halo. We note, however, that the division in two clumps is not obvious in either the diffuse X-ray emission or the infrared star counts. Indeed, \citet{brandner} place the centre of the cluster in a position very close to WR L, in the area of lower density of supergiants. This may suggest that the division in two clumps is not real, but due to a random decrease in the number of optically bright stars. In such case, the shape of the cluster, as defined by its supergiant (and WR) populations would resemble more closely an arc than an elongated structure.

Some objects, like W228b and W373, are located more than $150\arcsec$ away from the putative core centre, hinting at the possibility that Wd~1 is surrounded by a very extended halo. Though this possibility has not been explored yet, it is given strong support by the presence of 3 WR stars likely associated with the cluster at much higher distances \citep[WR stars N, X and T in the notation of ][]{wrs,crowther}. 

At this level of analysis, we find no strong indication of any difference in the relative distributions of evolved and unevolved stars, which could point to an age difference between regions inside Wd~1. The visually brightest stars seem to be concentrated in the small core defined above, but, given that neither the less evolved supergiants (their progenitors) or the WR stars (their descendants) share this concentration, this may be a random effect. In view of this, we conclude that the data so far strongly point to a single, homogeneous age (though some minimal spread, $\sim1$~Myr, is allowed) for the whole cluster.

\section{Conclusions}

Using new classification criteria, developed in Appendix~A, we provide spectral types for almost 60 OB supergiants in Wd~1. Only a small fraction of them had previous accurate classifications.

The stars leaving the cluster main sequence form a well populated clump at spectral types O9--B0, displaying a range of luminosities, which likely reflect a difference in initial rotational velocities and, perhaps, formation over a timespan. Sixteen luminous supergiants, with spectral types in the range B1--B4 range span the temperature range between this clump and the three blue hypergiants that bridge the gap with the A/F hypergiants. 

After a careful analysis of colour excesses, we find no strong reason to suspect that the extinction law towards Wd~1 deviates strongly from the standard, and we attribute the slightly divergent $E(B-V)$ values to the difficulty in obtaining precise $B$ photometry. The measured values of $E(V-I)$ and $E(V-R)$ show that the extinction varies strongly across the face of the cluster, with average values implying $A_{V}=10.8$. This value is fully consistent with the $A_{K_{{\rm S}}}=1.13\pm0.03$ derived by \citet{brandner} and a standard law. The value $A_{K_{{\rm S}}}=0.96$ found by \citet{crowther}, though subject to much higher uncertainty, is also compatible.

The dereddened magnitudes of the OB supergiants support a distance to Wd~1 similar to that found by \citet{crowther} from the infrared magnitudes of WR stars, namely $DM=13.5$ ($d=5.0$~kpc). With this distance modulus, the intrinsic magnitudes of OB supergiants correspond closely to their spectral types according to typical calibrations. The shorter $DM=13.0$ ($d=4.2$~kpc) found by \citet{kothes} is disfavoured (though their error bars reach $d=4.9$~kpc). The $DM=12.75$  ($d=3.5$~kpc) found by \citet{brandner} is strongly disfavoured, as it would mean that all our Ia supergiants would have intrinsic magnitudes typical of Iab supergiants, implying masses $\la25\,M_{\sun}$, very difficult to reconcile with their age derivation of 3.6~Myr. We suspect this to be due to the systematically younger ages that pre-MS isochrones indicate \citep[cf.][]{naylor}.

The picture drawn from the distribution in spectral types seems compatible with a single burst of star formation and the predictions of evolutionary models until the stars start to move quickly towards lower temperatures. Further studies, including the use of the population of unevolved stars to nail down the cluster age and detailed analysis of the evolved population (which will provide accurate stellar parameters and element abundances), will undoubtedly result in very firm constraints on the evolutionary tracks followed by massive stars after they leave the main sequence. The high binary fraction found amongst massive stars in Wd~1 \citep{clark08,ritchie09a} will likely allow a better understanding of the role that binarity plays in this evolution. Furthermore, characterisation of binary 
properties (frequency, mass ratio, period distribution, etc.) in the homogeneous massive star population of Wd~1 can provide strong constraints on the formation mechanism of massive stars in such dense environments\fnmsep\footnote{Note, however, that at an age $\sim5$~Myr, Wd~1 must have already undergone dynamical evolution. Indeed, \citet{brandner} find some evidence for mass segregation, which could partially erase the signature of the formation mechanism.}, as the  different formation mechanisms proposed will predict different 
outcomes \citep[for a review]{zy07}. Long-term spectroscopic monitoring of binaries in the cluster is underway \citep{ritchie09a}, and will deliver some of the observational characteristics.

\begin{acknowledgements}

We thank Lucy Hadfield for help with the 2004 observations and Dr. Amparo Marco for participating in other runs. We also thank Antonio Flor\'{\i}a for help with the artwork in the finder. Finally, we thank the referee, Micha\"el De Becker, for helpful comments.
This research has been funded by grants AYA2008-06166-C03-03 and
Consolider-GTC CSD-2006-00070 from the Spanish Ministerio de Ciencia e
Innovaci\'on (MICINN).  JSC acknowledges the support of an RCUK fellowship.

This research has made use of the Simbad database, operated at CDS,
Strasbourg (France).
\end{acknowledgements}

\appendix
\section{Spectral classification of OB supergiants in the red}

\subsection{Historical introduction}
OB stars have spectra characterised by strong Balmer and He
lines. Metallic lines are few and weak, except in the range
O9\,--\,B2, especially in main sequence stars. The blue range
concentrates most of the salient spectral features and spectral
classification has been developed using features in the
$\lambda\lambda$ 3900\,--\,4700 range \citep[e.g.,][]{wal71,waf}. 
The red part of the spectrum
contains fewer features and has generally been ignored, except for
H$\alpha$, considered a good tracer of mass loss. An overview of
features in the  $\lambda\lambda$ 5400\,–-\, 6600\AA\ range is
presented by \citet{wal80}, together with a spectral atlas based on
photographic spectrograms. High-quality digital spectrograms in this
range of a few selected objects are presented in \citet{wh00}. 

Further to the red, atmospheric features become a complication. There
is, however, an important window, free from atmospheric features, in
the  $\lambda\lambda$ 8300\,--\,8900 range. \citet{and95} present
spectra of early-type stars in the $\lambda\lambda$ 8375\,–-\,8770 range,
concluding that supergiants can be distinguished from main sequence
stars from the shape of the Paschen lines. According to these authors,
stars earlier than B3 cannot be classified in this region. Higher
resolution spectra were presented by \citet{mt99}. Their spectral
atlas covers the $\lambda\lambda$ 8480\,–-\,8750 range at $R=20\,000$,
but little attention is paid to early type stars. \citet{mt99} find
some degeneracy between luminosity class and rotational speed and
conclude that O-type stars cannot be classified at all with this
limited range.

\citet{caron} studied quantitative classification criteria, using
intermediate-resolution spectra in the $\lambda\lambda$ 8400\,-–\,8900
range for stars with spectral types between O9 and B5. They found that
the lines are always broader in dwarfs than in supergiants. In
addition, the equivalent widths of 
most lines increase with advancing spectral type. Based on these
findings, they developed a method for approximate classification based
on line ratios. This approach was followed in \citetalias{main}, where the grid
was extended by using synthetic spectra and the validity of the
classifications was checked against spectra of stars of known spectral
types (see Appendix A in \citetalias{main}).

Quantitative analysis of features in the red region have hardly been
reported. The only exceptions are \ion{C}{ii}~6578\AA,
\ion{Si}{ii}~6347\AA\ and \ion{Ne}{i}~6402\AA,
included by \citet{lennon93} in
their analysis of features seen in B-type supergiants. In addition, \citet{davies05} present an atlas of luminous supergiants in this range. 

\subsection{Data used}
In an attempt to improve our knowledge of the behaviour of features in
the red and far red regions, we have collected spectra of OB stars in
this range from a number of sources. The STELIB library
\citep{leborgne} contains stellar spectra at $R\sim 2000$. The data
cover the whole optical range, but the number of OB supergiants
observed is small. The library by \citet{cenarro} covers the
$\lambda\lambda$8350\,--\,8900 range at better spectral resolution,
but, being conceived to 
study the \ion{Ca}{ii} triplet, contains few early-type stars. The
more recent MILES library \citep{miles}, which includes more
early-type stars, only extends until 7500\AA. 

The main sources of data have been the Indo-US library \citep{indous},
which includes spectra in the range $\lambda\lambda$3460\,--\,9460 at
intermediate resolution and the UVES Paranal Observatory Project (ESO
DDT Program ID 266.D-5655) spectra \citep{bagnulo}. The UVES POP
spectra have excellent resolution $R\approx80\,000$ and
SNR. Unfortunately, as the data were taken as part of a poor weather
backup programme, the red sections are generally affected by very
strong atmospheric features. Moreover, the spectra present a gap in
the $\lambda\lambda$8540\,--\,8660 interval, due to the configurations
used. 

In all, the spectra represent a rather complete, if heterogeneous,
sample of B-type luminous stars. The O subtypes are more sparsely
sampled. 

\subsection{Results}

 In spite of the scarcity of spectral features in the red spectra of OB
stars, several criteria can be used for spectral
classification. Unfortunately, most of these criteria are 
sensitive to both spectral type and luminosity, but by considering the
whole spectrum in the $\lambda\lambda$6000\,--\,9000 range and
combining several criteria, we can narrow the spectral range for a given
supergiant spectrum generally down to half a spectral subtype. The
classification of dwarf stars is much less accurate. In what follows,
we will concentrate on the features used to classify the supergiants
seen in Wd~1.

\subsubsection{Classification in the $I$-band}

Criteria for the classification of O9\,--\,B5 stars in this range have
been laid down by \citet{caron}. They base their classification in the
relative strengths of Paschen lines. The best temperature criteria seem
to be the EW(Pa\,16+\ion{C}{iii})/EW(Pa\,11) or
EW(Pa\,15)/EW(Pa\,11) ratios, even though
\ion{He}{i}~8850\AA\ is blended with Pa\,11 at moderate
resolutions. The ratio FWHM(Pa\,13)/FWHM(\ion{He}{i}~8779\AA) is a good
luminosity criterion at the resolution they use. Unfortunately, in the
spectra of stars in Wd~1, suffering from heavy extinction, there seems 
to be a complex system of  diffuse interstellar bands. The 8620\AA\
band is sharp and 
narrow. \citet{munari08} has reported an excellent correlation between
its strength and reddening up to rather high reddenings, at least
$E(B-V)\sim2$. However, at the much higher reddening of Wd~1, this
feature is contaminated by neighbouring ones. There is a broad,
relatively shallow feature, perhaps similar in shape to the 4430\AA\ DIB,
centred around 8648\AA. This feature is blended with a
\ion{He}{i}~8653\AA\ line, weakly seen in many spectra of B-type stars
in \citet{caron}.

In addition, there appears to be a global
depression of the continuum in the whole range 8540\,--\,8660\AA. The
actual nature of this depression is difficult to decide, but it has a
strong effect on continuum normalisation. Treating it
as a single feature produces an awkward shape for the continuum and
results in a wrong sequence of strengths for the Paschen lines. In
view of this, for the purposes of this paper, we have normalised the
spectrum as if the continuum around 8610\AA\ were not affected by this
broad depression (effectively assuming that there are two such
depressions, one on each side of 8610\AA) and we have normalised the
spectra taking reference points in the continuum at 8520, 8610 and
8690\AA. This procedure results in 
a decrease of the strength of the Paschen lines as we move up the
sequence and it is hence taken to be approximately correct. However,
it is only an approximation and we assume that the equivalent widths
of Pa\,13, 14 and 15 cannot be measured with
accuracy. Because of this, we are forced to try to define a morphological classification scheme.

   \begin{figure}
   \centering \resizebox{\columnwidth}{!}{\includegraphics{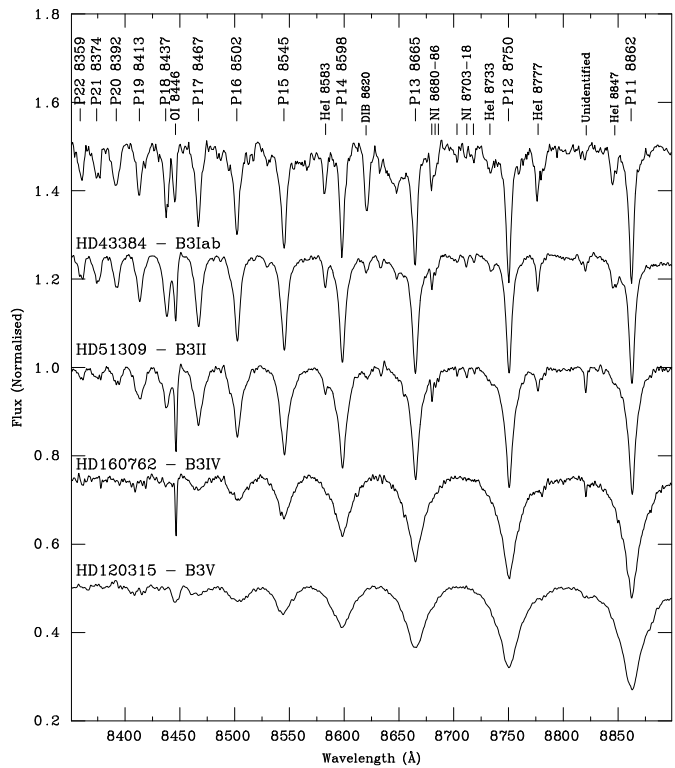}}
   \caption{Luminosity sequence at B3. The unlabelled spectrum at the
     top is W70 (B3\,Ia). All the other spectra have been taken from
     the Indo-US library. Note the strengthening and sharpening of
     \ion{He}{i} and \ion{N}{i} lines with increasing luminosity and
     the changing Pa~18/\ion{O}{i}~8446\AA\ ratio. The spectrum of W70
   is affected by the strong interstellar features mentioned in the text.}
              \label{figb3}%
    \end{figure}

Several \ion{He}{i} lines are seen in the $I$-band spectra of B-type
stars. Though never near as prominent as the Paschen lines, they are
much better defined in supergiants, showing maximum strength at B2\fnmsep\footnote{The \ion{He}{i} 8582.7 \& 8776.7 \AA\ lines were first described by \citet{helden72}, who discusses their sensitivity to luminosity}. In
dwarf stars, \ion{He}{i} lines are always weak and only seen in the
O8-B2 range (perhaps visible in high-SNR spectra of B3\,V slow
rotators). In giants, they are visible up to B4--5, depending on
$v_{{\rm rot}}$. In supergiants, they are seen down to B6 or even
later if the SNR is very good. 

   \begin{figure}
   \centering
 \resizebox{\columnwidth}{!}{\includegraphics{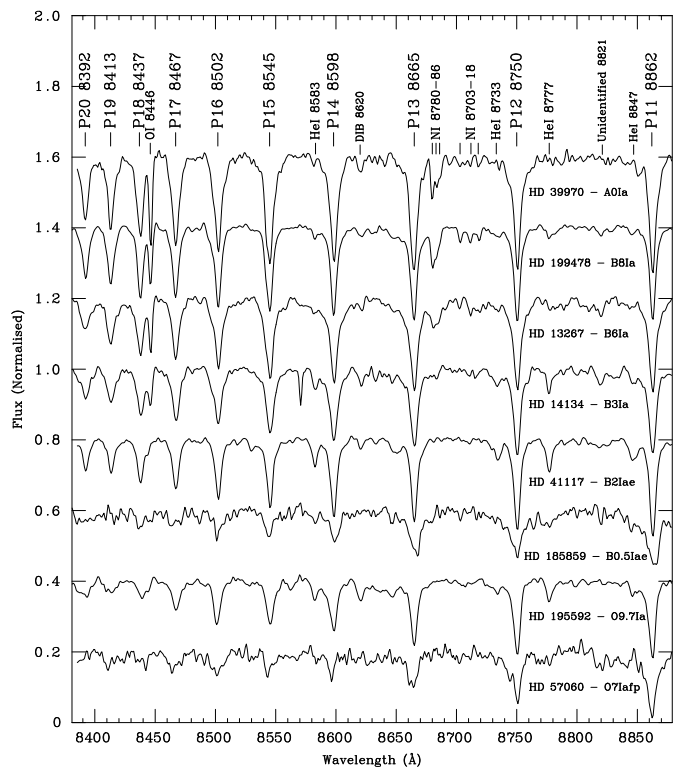}}
   \caption{Spectral sequence of high-luminosity supergiants. Spectra
     are taken from \citet{cenarro} and have only moderate resolution
     and occasionally low SNR.  Note the increase of  \ion{N}{i} lines
     with advancing spectral type and the sharp maximum of the
     \ion{He}{i} around B2. Note also that the lines marked above the
     top spectrum are those seen in B-type spectra and do not match
     those on the A0\,Ia spectrum.}
              \label{figlowres}%
    \end{figure}

The \ion{O}{i}~8446\AA\ line has a strong dependence on both spectral
type and luminosity. As shown in Fig.~\ref{figb3}, at a given spectral
type, the ratio with the neighbouring Pa~18 (8437\AA) decreases slowly
with increasing luminosity. Figure~\ref{figlowres}, on the other hand,
shows how the same ratio increases with spectral type at a given
luminosity. To make things worse, the shape of the line depends
strongly on the rotational velocity. This can be easily seen by comparing
the spectrum of HD~120315 (B3\,V, $v_{{\rm rot}}=150\: {\rm km}\,{\rm
  s}^{-1}$) and HD~160762 (B3\,IV and a very slow rotator; see
\citealt{abt02}). In any case, \ion{O}{i}~8446\AA\ is first seen at B2
in dwarf stars and starts to appear on the wing of Pa~18 at B1.5 in
supergiants. 

An important diagnostic for B3 is the appearance of \ion{N}{i} lines. They are weak at this spectral type, but quickly become stronger at later types, becoming prominent in supergiants after B5. These lines are not seen in dwarf stars. They are not seen in most spectra of luminosity class III objects, even the high-resolution UVES spectra. They are clearly visible in the few luminosity class II spectra available (including the B3\,II star in Fig.~\ref{figb3}), and so it is reasonable to expect that they may appear weakly in some giants with low  $v_{{\rm rot}}$. In fact, they are weakly visible in the spectrum from \citet{and95} of HD~186568 (B8\,III), which is a very slow rotator \citep[$v_{{\rm rot}}=15\: {\rm km}\,{\rm s}^{-1}$;][]{abt02}.

   \begin{figure}
   \centering
 \resizebox{\columnwidth}{!}{\includegraphics{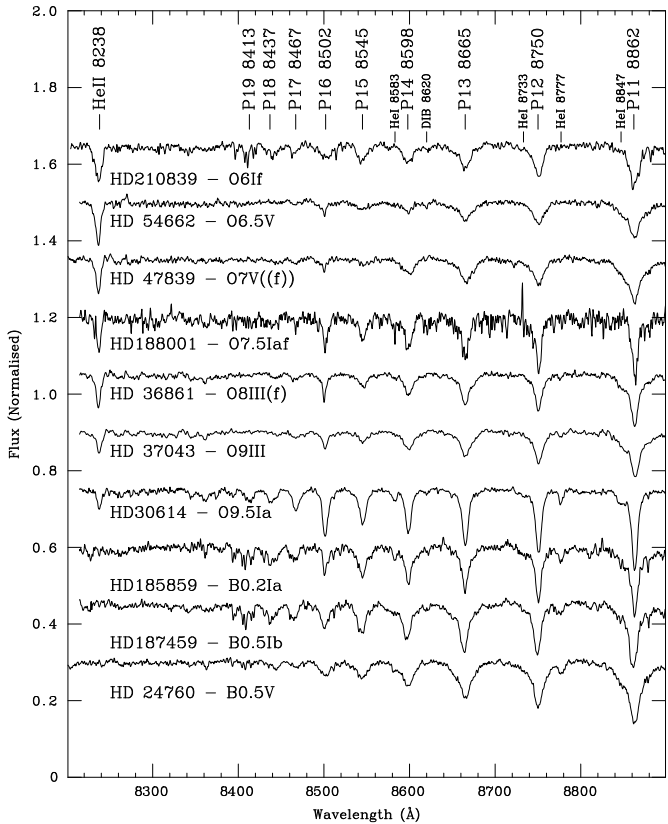}}
   \caption{$I$-band spectra of O- and early B-type stars. These
     spectra from the Indo-US library have had the telluric features
     removed and allow the study of the behaviour of
     \ion{He}{ii}~8238\AA. Note how this line clearly identifies
     O-type stars. Note also the much higher number of Paschen lines
     seen in supergiants at a given spectral type. Finally, the
     \ion{C}{iii}~8500\AA\ line is an excellent indicator of late
     O-type stars.}
              \label{figearlies}%
    \end{figure}

As discussed in \citetalias{main}, O-type stars can be easily identified by the
presence of \ion{He}{ii}~8238\AA. Unfortunately, this line lies in the
middle of a strong atmospheric band. When telluric features are
accurately removed, \ion{He}{ii}~8238\AA\ can be readily identified
(Fig.~\ref{figearlies}). Then it is visible up to at least O9.5, though
it may be difficult to locate in fast rotating dwarfs. When the
telluric features cannot be removed, as in the UVES POP spectra, it is
almost impossible to see, though it may be guessed in the spectra of
early and mid O-type stars. Based on the few spectra available, this
line may show some dependence on spectral type, being weaker in O8-9
stars. No strong dependence on luminosity class can be identified.

The second important marker of O-type stars is
\ion{C}{iii}~8500\AA. This line almost coincides with Pa\,16 (8502\AA) in
wavelength. In O6--7 dwarfs, where Pa\,16 is hardly seen, it appears as
a weak narrow line (Fig.~\ref{figearlies}). In O8-B0 stars it
can be noticed as a important strengthening of Pa\,16 in comparison with
Pa\,15 and Pa\,17. This strengthening stops being noticeable around
B0.5, in line with the behaviour of \ion{C}{iii} lines in the blue. As other C lines, this might be an unreliable tracer in
stars with CNO anomalies, though we lack appropriate spectra to test
this issue.

\subsubsection{Classification in the red region}

Spectral features in the red region are scarce in OB
stars. Many dwarfs show only H$\alpha$ and the \ion{He}{i}~6678 \& 7065\AA\ lines. Supergiants present a few other interesting features, notably in B-types.

\begin{figure}
   \centering
 \resizebox{\columnwidth}{!}{\includegraphics{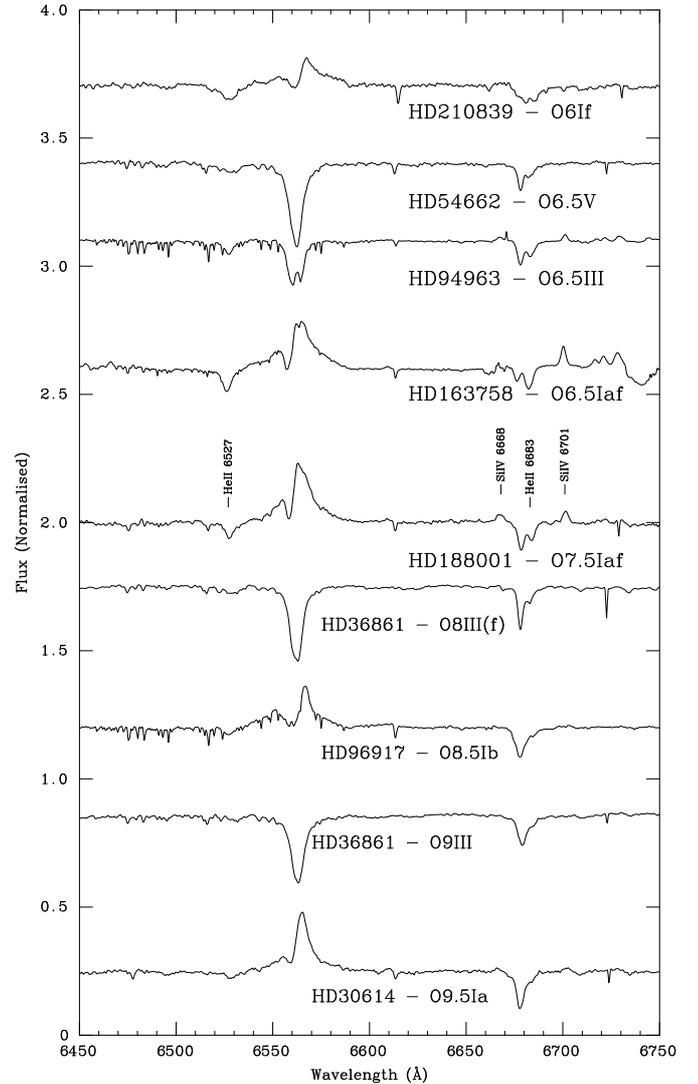}}
   \caption{Spectral features seen around H$\alpha$ in O-type
     stars. The spectra are drawn from several sources and have quite
     different resolutions. The \ion{He}{i}~6678\AA\ line has not been
   marked to prevent confusion.}
              \label{figoalpha}%
    \end{figure}

  There are two \ion{He}{ii} lines present in O-type
spectra (a third one, \ion{He}{ii}~6560\AA, cannot be separated from H$\alpha$). Unfortunately, \ion{He}{ii}~6527\AA\ coincides with a strong
DIB and cannot be used for classification. \ion{He}{ii}~6683\AA\
starts to be seen as a weak inflection on the wing of
\ion{He}{i}~6678\AA\ in high-SNR spectra of O9 stars, and it becomes
stronger at earlier spectral types. However, a clear dependence on
spectral type cannot be drawn from the few spectra available
(Fig.~\ref{figoalpha}) . The
ratio \ion{He}{i}~6678\AA/\ion{He}{ii}~6683\AA\ cannot be used to
determine spectral type. Indeed, from the four stars with spectral
type O6--6.5, it would seem to depend more strongly on luminosity.

A clear marker of high luminosity in O-type stars is the presence of
the \ion{Si}{IV}~6668,~6701\AA\ lines in emission. These are ``wind''
or selective emission
lines, due to non-LTE effects in an extended atmosphere. They are seen
in luminous supergiants earlier than O9. We lack information on their
behaviour in early O-types. The \ion{Si}{iv} lines in the blue part of
the spectrum go into emission at much earlier spectral types
\citep{walborn02}. 

%

   \begin{figure}
   \centering
 \resizebox{\columnwidth}{!}{\includegraphics{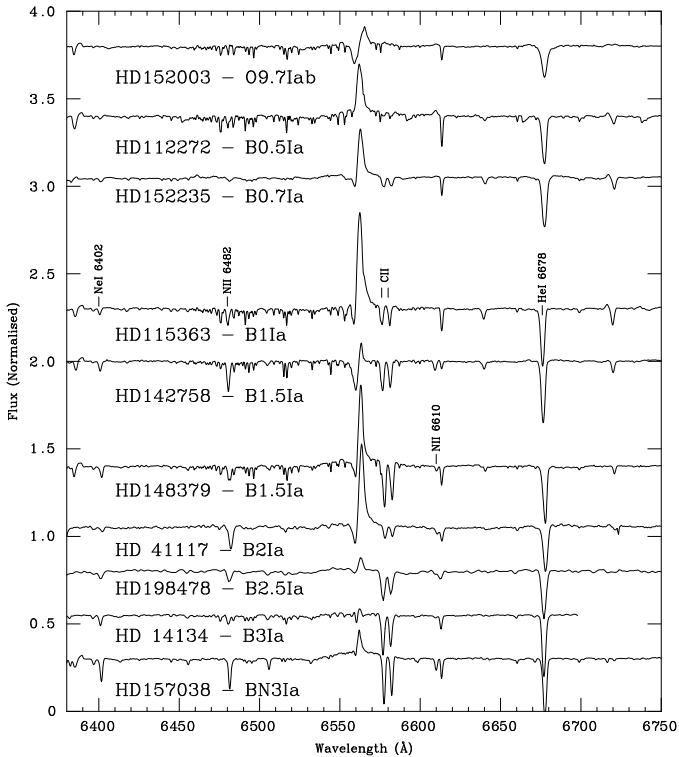}}
   \caption{Spectral features seen around H$\alpha$ in B-type supergiants
     stars. The spectra are drawn from several sources and have quite
     different resolutions. Spectra from the UVES POP show strong
     atmospheric features. The correlation
     between the intensity of the \ion{C}{ii}~6578, 6582\AA\ doublet
     and spectral type is loose because of CNO anomalies, as proved by
   the strong anticorrelation with the strength of \ion{N}{ii} lines
   at a given spectral type.}
              \label{figbalpha}%
    \end{figure}
Moving to later spectral types, we find another interesting ``wind''
feature, the \ion{C}{ii}~7231, 7236\AA\ doublet. Unfortunately, this
feature lies in the middle of an atmospheric band, but, being in
emission, is easily identifiable. This feature is visible between O9.7
and B2. Its strength is clearly correlated to that of H$\alpha$
emission, though there may be some dependence on spectral type, as
well, with stronger features being seen in the B0.5-1 range. It is
restricted to stars with heavy mass loss, and so generally only
appears in luminosity class Ia supergiants.

The behaviour of H$\alpha$ is similar. This line is a good tracer of
mass loss \citep{leitherer}. It tends to be in emission in luminosity
class Ia supergiants (as shown in Fig.~\ref{figbalpha}) and appears
normally in absorption in luminosity class Ib stars. Luminosity class
Iab supergiants occupy a middle ground, with H$\alpha$ weakly in
absorption or neutralised on many occasions, but in emission or strong
absorption in other cases. There are many exceptions to this general rule \citep[see more examples in][Appendix A]{davies05}. In many cases, variable H$\alpha$ emission is taken as evidence
for binarity \citep[e.g.,][]{thaller}, but monitoring of a few luminous OB supergiants has shown that important changes can happen on very short timescales \citep{morel04}. This behaviour has also been observed amongst the brightest OB supergiants in Wd~1 \citep{clark09}.

The shape of H$\alpha$ in emission has some dependence on spectral
type. In B-type supergiants, it appears as a P-Cygni profile,
sometimes superimposed on a broader plinth. The absorption trough
becomes weaker around B1-2 and tends to disappear completely. B0-0.5
supergiants generally show a strong single peak. O-type supergiants
tend to have more complex features, with more than one peak. This is
complicated by the presence of the \ion{He}{ii}~6560\AA\ line in
absorption. 

Emission is weaker for late-B supergiants, which generally have weak
P-Cygni profiles, but there are also many exceptions. For example, the
B8\,Ia supergiant HD~199478 can show a strong 
H$\alpha$ single peak emission, though this line is variable
\citep{mv00}. The most luminous hypergiants may also display strong
symmetric peaks. For example HD~32034 (B8-9\,Ia$^{+}$) or HD~269781
(A2\,Iae) in the catalogue of \citet{leborgne} do, but HD~33579
(A3\,Ia$^{+}$) and HD~223385 (A3\,Ia$^{+}$) display a weak absorption
trough superimposed on the strong emission peak (Note that the
extremely high luminosity of all these stars, except HD~223385, can be
confirmed, as they are members of the LMC). 

The \ion{O}{i}~7774\AA\ triplet is a very good luminosity
indicator. Calibrations of its strength against luminosity exist for
A\,--\,G stars \citep[e.g.,][]{arel03}, but
its behaviour has not been tabulated in B-type stars. The lines are
barely visible in B1 supergiants at moderate resolution, but quickly
grow in strength with advancing spectral type. They are seen in dwarfs
from B2--3, depending on rotational velocity.


The \ion{C}{ii}~6578, 6582\AA\ doublet is generally not seen in MS B-type stars, where it is very weak and easily masked by high rotation or low SNR. It is present in B2--B5 stars of moderate luminosity and in most B-type
supergiants \citep{wal80}. As it is clearly seen in many luminosity class IV stars, we suspect that it may be seen in MS stars with very low $v_{{\rm rot}}$. At high SNR, it is weakly visible in B0.5
supergiants. It clearly appears in B0.7 supergiants and its strength
peaks around B3 \citep{lennon93}. The lines are visible until B8, and also in some B9 supergiants, with a clear dependance on luminosity. As seen in Fig.~\ref{figbalpha}, their strength is not monotonically correlated to spectral type amongst luminosity class Ia supergiants. There seems to be some degree of anticorrelation between the strength of 
these features and those of \ion{N}{ii}, suggesting that differences
in CNO abundances play a role in their relative strength, as is seen
in features further to the blue \citep{wh00}.

  \begin{table*}
\label{features}
\begin{center}
\caption{Summary of features useful for spectral classification between O4 and B9 in the 6300--9000\AA\ range.}
\begin{tabular}{lllll}
\hline
Feature & from & to & maximum& Comments\\ 
     &  &            &             & \\     
\hline
Paschen lines   &  O4 & B9 & $>$B9 & Number of lines seen depends on temperature and luminosity\\
& & & & Shape of lines correlates with luminosity\\ 
\ion{He}{i} 6678, 7065\AA\ & O4 & $\sim$B7\,V, B9\,I & B2 & \\
\ion{He}{i} 8847, 8777, 8583\AA\ &O8--9 & B3\,V, B6\,I& B2 & strength depends on luminosity \\
\ion{He}{ii} 6683\AA\ & O4 & O7\,V, O9\,I & O4 & blended with \ion{He}{i} 6678\AA \\
\ion{He}{ii} 8283\AA & O4 & O9\,I--V & O4 & inside telluric band\\
\ion{C}{iii} 8500\AA & $\sim$O6 & B0.5 & O9 & lower limit not well determined\\
\ion{C}{ii} 6578, 6582\AA & B0.7 & B6\,III, B8\,I & B3 & rarely seen in dwarfs\\
\ion{C}{ii} 7231, 7236\AA & O9.7\,I & B2\,I & B1 & emission features only in luminous SGs\\
&&&& inside telluric band\\
\ion{O}{i} 8446\AA & B2\,V B1.5\,I & B9 & $>$B9 & strong dependence on luminosity\\
\ion{O}{i} 7774\AA & B3\,V B1\,I & B9 & $>$B9 & strong dependence on luminosity\\
\ion{Si}{ii} 6347, 6371\AA & B2\,I & B9 & B8 & tabulated by \citet{lennon93} in SGs\\
\ion{Si}{iv} 6668, 6701\AA & $-$ & O9 & ? & emission lines in SGs; lower limit not determined\\
\ion{N}{i} 8680--86, 8703--18\AA & B3\,I--II & B9\,I-II &$>$B9  & not seen in III--V\\
\ion{N}{ii} 6482, 6610\AA & B1\,I & B3\,I & B2  & only strong in Ia SGs\\
\ion{Ne}{i} 6402\AA & B1\,I & B9\,I & B4--5  & tabulated by \citet{lennon93} in SGs\\
\hline
\end{tabular}
\end{center}
\end{table*}

%

There are two \ion{N}{ii} lines in the range showed in
Fig.~\ref{figbalpha}, which are only visible 
in supergiants. \ion{N}{ii}~6482\AA\ appears weakly in B1 supergiants,
is rather strong at B2 and becomes weak again at B3, disappearing at
later spectral types. Its dependence on temperature is very similar to that of
\ion{N}{ii} lines in the blue range, such as
\ion{N}{ii}~4601\AA, tabulated by
\citet{lennon93}. \ion{N}{ii}~6610\AA\ is rather weaker, and has the
same temperature dependence. As shown by \citet{lenn93}, these two
lines are only prominent in B2\,Ia supergiants. However, N-enhanced
stars of similar spectral types may also show relatively strong
features (Fig.~\ref{figbalpha}). Another \ion{N}{ii} line behaving
similarly is \ion{N}{ii}~6380\AA\ \citep{wal80}.

The \ion{Ne}{i}~6402\AA\ line is seen in several
of the spectra. Its behaviour was tabulated by \citet{lennon93}. The line is
visible from B1 to B9, and strongest at B4--5, but it does not seen to
depend on luminosity subclass amongst supergiants. A weaker line,  \ion{Ne}{i}~6506\AA, is seen in the spectrum of HR~Car, which is classified B2 \citep{nota97}. The
\ion{Si}{ii}~6347\AA\ line was also tabulated by \citet{lennon93}. It
is hardly visible at B2, but increases its strength very quickly,
becoming prominent at B3 and growing in strength towards later
spectral types. The neighbouring \ion{Si}{ii}~6371\AA\ behaves
    similarly. Features further to the blue are discussed by \citet{wal80}. 

\subsection{Summary and outlook}
The results presented here show that classification of early-type stars is possible using only red spectra. A summary of features useful for classification is presented in Table~A.1. The classifications are not as accurate as those using blue spectra, but likely comparable to classifications based on $K$-band spectra. Combination of red spectra with near-IR spectra offers good prospects to classify rather accurately obscured early-type stars. The limited spectral range offered by GAIA does not allow accurate spectral classification, but the situation is not as desperate as suggested by \citet{mt99}. As seen in Fig.~\ref{giants}, even O-type stars can be approximately classified in that range, at high resolution and signal-to-noise ratio.

Further improvement on classification and a more thorough understanding of some of the issues raised here may be obtained by a systematic study of MK standards and stars with accurate spectral types.


\begin{thebibliography}{}

\bibitem[Abt et al.(2002)]{abt02} Abt, H.A., Levato, H., \& Grosso,
  M. 2002, ApJ, 573, 359

\bibitem[Andrillat et al.(1995)]{and95} Andrillat, Y., Jaschek,
  C., \& Jaschek, M. 1988, A\&AS, 112, 475

\bibitem[Arellano Ferro et al.(2003)]{arel03} Arellano Ferro A.,
  Giridhar S. \& Rojo Arellano E., 2003, RMxAA, 39, 3  

\bibitem[Bagnulo et al.(2003)]{bagnulo} Bagnulo, S., Cabanac, R.,
  Jehin, E., et al. 2003, Messenger, 114, 10

\bibitem[Bonanos(2007)]{bonanos} Bonanos, A.Z. 2007, AJ, 133, 2696

\bibitem[Brandner et al.(2008)]{brandner} Brandner, W., Clark, J.S.,
  Stolte, A., et al. 2008, A\&A, 478, 137

\bibitem[Caron et al.(2003)]{caron}
Caron G., Moffat A.F.J., St-Louis N., et al. 2003, AJ, 126, 1415

\bibitem[Clark \& Negueruela(2004)]{cn04}
Clark J.S., Negueruela I., 2004, A\&A, 413, L15

\bibitem[Clark et al.(2005)]{main} 
Clark, J.S., Negueruela, I., Crowther, P.A., \& Goodwin, S.P. 2005,
A\&A, 434, 949 (C05)

\bibitem[Clark et al.(2008)]{clark08} 
Clark, J.S., Muno, M.P., Negueruela, I., et al. 2008, A\&A, 477, 147 

\bibitem[Clark et al.(2010)]{clark09} 
Clark, J.S., Ritchie, B.W., \& Negueruela, I. 2010, A\&A, in press

\bibitem[Cenarro et al.(2001)]{cenarro} Cenarro, A.J., Cardiel, N.,
  Gorgas, J., et al. 2001, MNRAS, 326, 959

\bibitem[Cox et al.(2005)]{cox05}Cox, N.L.J., Kaper, L., Foing, B.H., et al. 2005, A\&A, 438, 187

\bibitem[Crowther et al.(2006a)]{crowther}Crowther, P.A., Hadfield,
  L.J., Clark, J.S., et al. 2006a, MNRAS, 372,
  1407 

\bibitem[Crowther et al.(2006b)]{crowtherbsgs}Crowther, P.A., Lennon, D.J., \& Walborn, N.R. 2006b, A\&A, 446, 279 

\bibitem[Davies et al.(2005)]{davies05} Davies, B., Oudmaijer, R.D., \& Vink, J.S. 2005, A\&A, 439, 1107
    
\bibitem[Dougherty et al.(2010)]{dougherty} Dougherty, S.M., Clark, J.S., Negueruela, I., et al. 2010, A\&A, 58, A54

\bibitem[Draper et al.(2000)]{draper} Draper, P.W., Taylor, M., \&
  Allan, A. 2000, Starlink User Note 139.12, R.A.L.

\bibitem[Eldridge et al.(2008)]{eldridge08} Eldridge, J.J., Izzard, R.G., Tout, C.A. 2008, MNRAS, 384, 1109

\bibitem[van Genderen(2001)]{vGen01} van Genderen, A.M. 2001,
  A\&A, 366, 508 

\bibitem[Howarth et al.(1998)]{howarth}Howarth, I., Murray, J.,
  Mills, D., \& Berry, D.S. 1998, Starlink User Note 50.21, R.A.L.

\bibitem[Humphreys \& Davidson(1994)]{hd94} Humphreys, R.M., \&
  Davidson, K. 1994, PASP 106, 1025  

\bibitem[Humphreys \& McElroy(1984)]{hme84} Humphreys, R.M., \& McElroy, D.B. 1984, ApJ, 284, 565 

\bibitem[Kothes \& Dougherty(2007)]{kothes} Kothes, R., \& Dougherty,
  S.M. 2007, A\&A, 468, 993

\bibitem[Le Borgne et al.(2003)]{leborgne} Le Borgne, J.-F., Bruzual,
  G., Pell\'o, R., et al. 2003, A\&A, 402, 433

\bibitem[Leitherer(1988)]{leitherer} Leitherer, C. 1988, ApJ, 326, 356

\bibitem[Lennon(1993)]{lenn93} Lennon, D.J. 1993, Space Sci. Rev. 66, 127

\bibitem[Lennon et al.(1993)]{lennon93} Lennon, D.J., Dufton, P.L., \&
  Fitzsimmons, A. 1993, A\&AS, 97, 595

\bibitem[Markova \& Puls(2008)]{markova} Markova, N., \& Puls,
  J. 2008, A\&A, 478, 823

\bibitem[Markova \& Valchev(2000)]{mv00} Markova, N., \& Valchev,
  T. 2000, A\&A, 363, 995

\bibitem[Martins et al.(2005)]{martins}Martins, F., Schaerer, D., \&
Hillier, J. 2005, A\&A, 436, 1049  

\bibitem[Massey(2003)]{mas03}Massey, P. 2003, ARA\&A, 41, 15

\bibitem[McErlean et al.(1999)]{mcerlean} McErlean, N.D., Lennon, D.J., \& Dufton, P.L. 1999, A\&A, 349, 553

\bibitem[Mengel \& Tacconi-Garman (2008)]{mt07} Mengel, S., \&
  Tacconi-Garman, L.E. 2009, Ap\&SS, 324, 321

\bibitem[Meynet \& Maeder(2000)]{mm}
Meynet, G., \& Maeder, A. 2000, A\&A, 361, 101

\bibitem[Morel et al.(2004)]{morel04} Morel, T., Marchenko, S.V., Pati, A.K., et al. 2004, MNRAS, 351, 552

\bibitem[Munari et al.(2008)]{munari08} Munari, U., Tomasella, L., Fiorucci, M., et al. 2008, A\&A, 488, 969

\bibitem[Munari \& Tomasella(1999)]{mt99} Munari, U., \& Tomasella,
  L. 1999, A\&AS, 137, 521

\bibitem[Munari \& Zwitter(1997)]{mz97} Munari, U., \& Zwitter, T. 1997, A\&A, 318, 269

\bibitem[Muno et al.(2006)]{muno06} Muno, M.P., Law, C., Clark, J.S., et al. 2006, ApJ, 650, 203

\bibitem[Naylor(2009)]{naylor} Naylor, T. 2009, MNRAS, 399, 432

\bibitem[Negueruela \& Clark(2005)]{wrs} 
Negueruela, I., \& Clark, J.S. 2005, A\&A, 436, 541 

\bibitem[Nota et al.(1997)]{nota97}
Nota, A., Smith, L., Pasquali, A., et al. 1997, ApJ, 486, 338

\bibitem[Rieke \& Lebofsky(1985)]{rieke} Rieke, G.H., \& Lebofsky, M.J. 1985, ApJ, 288, 618

\bibitem[Ritchie et al.(2009a)]{ritchie09a} Ritchie, B.W., Clark, J.S., Negueruela, I., \& Crowther, P.A. 2009a, A\&A, 507, 1585

\bibitem[Ritchie et al.(2009b)]{ritchie09b} Ritchie, B.W., Clark, J.S., Negueruela, I., \& Najarro, F. 2009b, A\&A, 507, 1597

\bibitem[Salasnich et al.(1999)]{sal99} Salasnich, B., Bressan, A., \& Chiosi, C. 1999, A\&A, 342, 131

\bibitem[S\'anchez-Bl\'azquez et
  al.(2006)]{miles}S\'anchez-Bl\'azquez, P., Peletier, R.F.,
  Jim\'enez-Vicente, J., et al. 2006, MNRAS, 371, 703

\bibitem[Shortridge et al.(1997)]{shortridge}Shortridge, K., Meyerdicks, H., 
Currie, M., et al. 1997, Starlink User Note 86.15, R.A.L.

\bibitem[Smith et al.(2004)]{smith04} Smith, N., Vink, J.S., \& de Koter, A. 2004, ApJ, 615, 475 

\bibitem[Thaller(1997)]{thaller} Thaller, M.L. 1997, ApJ, 487, 380

\bibitem[Valdes et al.(2004)]{indous} Valdes, F., Gupta, R., Rose,
  J.A., et al. 2004, ApJS, 152, 251 

\bibitem[van der Hucht et al.(2003)]{lanza} van der Hucht, K.,
  Herrero, A., \& Esteban, C., eds., 2003, A Massive Star Odyssey:
  From Main Sequence to Supernova. Proceedings of IAU Symposium \#212,
  held 24--28 June 2001 in Lanzarote, Canary Islands, Spain. San
  Francisco: Astronomical Society of the Pacific 

\bibitem[van Helden(1972)]{helden72} van Helden, R. 1972, A\&A, 19, 388

\bibitem[Vanbeveren et al.(2007)]{vanbeveren07} Vanbeveren, D., Van Bever, J., \& Belkus, H. 2007, ApJL, 662, L107

\bibitem[Walborn(1971)]{wal71}Walborn, N.R. 1971, ApJS, 23, 257

\bibitem[Walborn(1980)]{wal80} Walborn, N.R. 1980, ApJS 44, 535

\bibitem[Walborn \& Fitzpatrick(1990)]{waf} Walborn, N.R., \&
  Fitzpatrick, E.L. 1990, PASP, 102, 379

\bibitem[Walborn \& Howarth(2000)]{wh00} Walborn, N.R., \&
  Howarth, I.D. 2000, PASP, 112, 1446

\bibitem[Walborn et al.(2002)]{walborn02} Walborn, N.R.,
  Howarth, I.D., Lennon, D.J., et al. 2002, AJ, 123, 2754

\bibitem[Wegner(1994)]{wegner} Wegner, W. 1994, MNRAS, 270, 229

\bibitem[Westerlund(1987)]{westerlund}Westerlund, B.E. 1987, A\&AS, 70, 311

\bibitem[Zinnecker \& Yorke(2007)]{zy07} Zinnecker, H., \& Yorke, H.W. 2007, ARA\&A, 45, 481

\end{thebibliography}
\end{document}